# Evidence for a tricritical point coinciding with the triple point in $(Pb_{0.94}Sr_{0.06})(Zr_xTi_{1-x})O_3$ : A combined synchrotron X-ray diffraction, dielectric and Landau theory study

Ravindra Singh Solanki, [1] S. K. Mishra, [2] Yoshihiro Kuroiwa, [3] Chikako Moriyoshi[3] and Dhananjai Pandey[1]

[1]School of Materials Science and Technology, Indian Institute of Technology (Banaras Hindu University), Varanasi-221005, India

[2]Research and Technology Development Centre, Sharda University, Greater Noida-201306, India

[3]Department of Physical Science, Graduate School of Science, Hiroshima University, Japan

**Abstract**

We present here results of high resolution synchrotron X-ray diffraction (SXRD) and dielectric studies in conjunction with Landau theory considerations on $(Pb_{0.94}Sr_{0.06})(Zr_xTi_{1-x})O_3$ compositions in the vicinity of the morphotropic phase boundary (MPB) to find evidence for the flattening of the free energy surface at the MPB proposed in recent ab-initio studies on strongly piezoelectric ceramics. SXRD results reveal that the tetragonal and pseudorhombohedral monoclinic compositions with x=0.515 and 0.550 transform directly into the cubic paraelectric phase, whereas for $0.520 \leq x \leq 0.545$, the pseudotetragonal as well as pseudorhombohedral monoclinic compositions first transform to the tetragonal phase and then to the cubic phase. Our results reveal the existence of a triple point at $x \simeq 0.550$. It is shown that the tetragonal to cubic transition, irrespective of the composition, upto $x \simeq 0.545$ is accompanied with a discontinuous change in the unit cell volume and thermal hysteresis confirming first order nature of this transition. The pseudorhombohedral monoclinic composition for x=0.550, on the other hand, transforms directly into the cubic phase in a second order manner. Our experimental results thus reveal a crossover from first order to second order phase transition through a tricritical point around x=0.550. Landau theory calculations also confirm gradual flattening of the free energy surface on approaching the tricritical composition x=0.550. We conclude that the triple point in the PZT phase diagram is indeed a tricritical point.

## I. INTRODUCTION

Lead Zirconate Titanate, $Pb(Zr_xTi_{1-x})O_3$ (abbreviated as PZT), is the most widely used piezoelectric ceramic material in electromechanical sensor and actuator devices[1]. Its x-T phase diagram contains a morphotropic phase boundary (MPB) at $x \doteq 0.520$ that separates the stability fields of pseudorhombohedral monoclinic and tetragonal phases, stable for $x \geq 0.530$ and for $x < 0.520$, respectively, through a thin stability region of a bridging pseudotetragonal monoclinic phase[2-3]. The physical properties, like dielectric constant, piezoelectric constants and electromechanical coupling coefficients, show maximum response corresponding to the MPB composition. The ground state of PZT in the MPB region and the pseudorhombohedral monoclinic Zr-rich region has been a subject of controversy[4-5] but recent investigations suggest that the space group of the ground state of PZT near the MPB[6] and on $Zr^{4+}$ rich side of MPB is Cc[7]. Like the true ground state of PZT, the origin of high piezoelectric response for the MPB composition is also under intense debate[8-12] in recent years.

In a pioneering work, Fu and Cohen[8], using first principle calculations, have shown that the giant piezoelectric response of MPB ceramics is a consequence of flattening of the energy surface near the MPB which can facilitate the rotation of the polarization direction from $[001]_{pc}$ of the tetragonal phase towards the $[111]_{pc}$ of the rhombohedral phase on a symmetry plane of intermediate bridging phase(s) in the monoclinic symmetry. Such monoclinic phases have since been discovered and are known to possess Cm ($M_A$)[13], Cm ($M_B$)[14], Pm ($M_C$)[15] and Cc[2,6,7,16] space groups in different Pb-based MPB systems. Recently[17], in a lead free $Ba(Zr_{0.2}Ti_{0.8})O_3$-$x(Ba_{0.7}Ca_{0.3})TiO_3$ system, it has been argued that the physical properties show maximum response in the MPB region due to the existence of a tricritical point at the triple point where two first order phase boundaries (i.e., MPB and ferroelectric-paraelectric) intersect. The presence of a tricritical point leads to a nearly vanishing polarization anisotropy (flatter free energy surface) and can thus facilitate polarization rotation between tetragonal and rhombohedral states. The free energy flattening arising at high temperatures due to the presence of a tricritical point at the triple point can also result in flatter free energy surface at room temperature, leading to high piezoelectric response by means of polarization rotation for the MPB composition[11].

Not withstanding the significance of a tricritical point coinciding with the triple point of MPB ceramics in ensuring flattening of the free energy surface, there is considerable controversy regarding the evidence and location of tricritical point(s) in PZT. Based on dielectric and laboratory X-ray diffraction (XRD) data, Mishra et al[18], way back in 1997, argued that the triple point in the PZT phase diagram is a tricritical point. The possibility of the triple point in PZT being a tricritical point has been subsequently supported by Damjanovic[11] and Kim et al[19] also. Based on x-ray diffraction studies, Eremkin et al[20], on the other hand, have proposed the existence of two tricritical points at x= 0.78 and x = 0.45. Based on a theoretical analysis of the dielectric data and Landau theory considerations, Noheda et al[21] and Haun et al[22] have also proposed two tricritical points in PZT but at x = 0.74 and x = 0.49 and x = 0.898 and x= 0.283, respectively. Whatmore et al[23] using dielectric constant, remanant polarisation and spontaneous strain measurements found evidence for a tricritical point at the ferroelectric ($F_R$)-paraelectric ($P_c$) transition for x=0.94. A tricritical point on the cubic-to-tetragonal line of phase transition has been reported by Rossetti et al[24] at x=0.38 by calorimetric measurements. More recently, Rossetti et al[25] and Porta et al[12] using Landau theory considerations have proposed two tricritical points for PZT, one on the tetragonal side and another on the pseudorhombohedral side at x=0.30 and 0.80, respectively, across the triple point composition. The presence of two tricritical points implies that the nature of the paraelectric to ferroelectric phase transition is second order for all the compositions between the two tricritical points including the MPB which lies in between the two points. Since a second order phase transition can give rise to larger response functions than first order phase transition due to a bigger flattening of the free energy surface, it has been argued that this may hold clue to high piezoelectric response of PZT in the MPB region. However, it is not a priori clear from this model as to why the piezoelectric response should peak at the MPB composition only and not for the entire composition range between the two tricritical points.

The present work was undertaken to settle the existing controversies about the existence of a tricritical point near the MPB by analysing the nature of high temperature cubic to tetragonal and cubic to monoclinic phase transitions using temperature dependent dielectric and synchrotron X-ray powder diffraction studies across the MPB compositions of PZT with 6% Sr doping. We also use Landau theory considerations to study the flattening of the free energy surface on approaching the MPB from the tetragonal side. Our results present unambiguous evidence for the existence of a tricritical point at $x \simeq 0.550$ coinciding with the triple point. We have chosen 6% $Sr^{2+}$ doped PZT samples because they show enhanced piezoelectric properties and the ground state of such compositions in the MPB region has been settled unambiguously recently[5, 6].

## II. EXPERIMENT

Highly homogeneous samples of $(Pb_{0.94}Sr_{0.06})(Zr_xTi_{1-x})O_3$ (PSZT) with x=0.515, 0.520, 0.525, 0.530, 0.535, 0.545 and 0.550 were prepared by a semi-wet route which is known to give narrowest composition width ($\Delta x \approx 0.01$) of the MPB region in PZT[26] and PSZT[6(a)]. The details of PSZT sample preparation are given in our previous work[6(a)]. High resolution Synchrotron X-ray powder diffraction (SXRD) measurements as a function of temperature were carried out in the 100 to 800K range at BL02B2 beam line of SPring-8, Japan at a wavelength of 0.412Å (30 keV)[27]. The dielectric measurements were performed using a Schlumberger SI-1260 impedance gain phase analyser. For SXRD measurements as a function of temperature in the range 100-800K, the temperature was controlled to within ± $1^0$C. All the SXRD measurements were carried out in the heating cycle. The dielectric measurements were carried out during both heating and cooling cycles to study the thermal hysteresis. In these measurements, the sample was heated at the rate of $1^0$C/min but cooled at a much slower rate ($0.25^0$C/min) to capture genuine thermal hysteresis. The data were analysed by Rietveld technique using the FULLPROF[28] software package.

## III. RESULTS AND DISCUSSION

### A. Dielectric studies

Fig. 1 depicts the variation of dielectric constant as a function of temperature for various PSZT compositions. The peak in the dielectric constant at the Curie point of the sample occurs at a higher ($T_c^H$) value during heating cycle than that during the cooling ($T_c^C$) down of the sample. Such a thermal hysteresis in the heating and cooling cycles of the dielectric measurements is known to be due to the metastability of paraelectric and ferroelectric phases below and above $T_c$, in the cooling and heating cycles, respectively, due to a first order phase transition[29]. Fig. 2(a) shows the variation of $T_c$ during heating and cooling cycles with composition. The thermal hysteresis $\Delta T= T_c^H- T_c^C$ is ~9K for x=0.515 and decreases with increasing x. The extrapolation of the fitted curves, for both the heating and cooling cycles in Fig. 2(a) yields ΔT=0 for a composition x≃0.553. This shows that the nature of phase transition changes from first order with hysteresis to second order without hysteresis at x≃0.553 suggesting the possibility of a tricritical point at x≃0.553. Structural studies to be described in the next section were carried out to verify if the tricritical point at x≃0.553 coincides with the triple point of the PSZT phase diagram.

**B. Structural studies**

Pure single phase tetragonal and pseudorhombohedral monoclinic structures of PSZT are stable for x≤0.515 and x≥0.545, respectively[6(a)] in our chemically homogeneous samples. In the composition range 0.515<x≤0.525, a pseudotetragonal monoclinic phase coexists with the tetragonal phase, while the pseudorhombohedral monoclinic phase coexists with the pseudotetragonal monoclinic phase for 0.525<x≤0.535[6(a)]. The structure of the majority phase of PSZT stable across the MPB at room temperature is tetragonal (P4mm) for x≤0.525 and pseudorhombohedral monoclinic (Cm) for x≥0.545 with the intermediate pseudotetragonal monoclinic (Cm) phase for x=0.530 (MPB composition) and 0.535 which can act as a bridging phase[2, 30] between the tetragonal and pseudorhombohedral phases. With this background information about the room temperarature structures, we now proceed to analyse the high temperature behaviour of the various PSZT compositions.

Fig. 3 depicts the evolution of the $(111)_{pc}$, $(200)_{pc}$, and $(220)_{pc}$ (pc stands for pseudocubic) peaks of PSZT with x=0.515 (abbreviated as PSZT515) in the temperature range 100-800K. It is evident from this figure that at and above 625K, all the profiles are singlet. Below 625 K, the (200) and (220) cubic peaks split into two peaks, while (111) remains a singlet. These are the characteristics of a tetragonal phase. We therefore conclude that the cubic phase for x=0.515 transforms directly into the tetragonal phase below 625K.

For the PSZT compositions with x=0.520 and 0.525 (abbreviated as PSZT520 and PSZT525, respectively), the dominant phase at room temperature and below is tetragonal but a pseudotetragonal monoclinic phase coexists with it. This is evidenced by the asymmetric broadening of the $(111)_{pc}$ and the triplet character of the $(220)_{pc}$ peaks which are singlet and doublet for the pure tetragonal phase as can be seen from Figs. 4 and 5 which depict the evolution of SXRD profiles from 100 to 800K. It is evident from Fig.4 that the triplet nature of the $(220)_{pc}$ peak changes to doublet around 250K. However, there is additional broadening in $(111)_{pc}$ peak profile which persists up to 525K, as shown in Fig. 6 (a). This broadening indicates the coexistence of the monoclinic phase for which $(111)_{pc}$ peak is not a singlet unlike for the tetragonal phase for which it is a singlet. The disappearance of this broadening at ~550K suggests that the pseudotetragonal monoclinic phase has disappeared above 525K leaving behind pure tetragonal phase only as evidenced by the additional broadening of the $(200)_{pc}$ peak. To reliably determine the tetragonal to cubic phase transition temperature, we plot in Fig. 6(b) the full width at half maximum (FWHM) of the $(200)_{pc}$ peak as function of temperature. The fact that the FWHM of the $(200)_{pc}$ peak remains constant for T≥625K indicates that tetragonal phase has transformed to cubic phase at T≃625K. This suggest that the pure tetragonal phase is stable in the 550K≤T≤625K. In Fig. 6(b), we have only two data points in the tetragonal region because for the third point corresponding to T=550K, $(200)_{pc}$ is clearly splitted into two peaks. So this data point is not included in Fig. 6(b)). For PSZT525 also, the triplet nature of the $(220)_{pc}$ peak changes to doublet like around 325K (see Fig.5) However, the variation of the FWHM of $(200)_{pc}$ and $(111)_{pc}$ peak profiles with temperature shown in Fig. 7 reveals that the structure becomes pure tetragonal at T≳550K and pure cubic above T≳625K. Thus the sequence of phase transitions above room temperature for PSZT520 and PSZT525 is pseudotetragonal monoclinic (Cm) + tetragonal (P4mm) to pure tetragonal and then tetragonal to cubic ($Pm\bar{3}m$).

Rietveld refinements were carried out to confirm the conclusions arrived at from the analysis of the FWHM of $(111)_{pc}$ and $(200)_{pc}$ peaks. We briefly discuss the Rietveld refinement results for PSZT520; similar refinements were carried out for PSZT525 as well. Fig. 8 depicts the Rietveld fits for some selected perovskite reflections of PSZT520 at four representative temperatures for various plausible structural models. Coexistence of tetragonal and monoclinic phase (P4mm+Cm) gives better Rietveld fit with lower $\chi^2$ value as compared to that for pure monoclinic phase at 100K. Pure tetragonal phase model is ruled out at 100K as the $(111)_{pc}$ peak is obviously not a singlet. This situation persist upto 225K. In the temperature range 225K<T<500K, the triplet nature of $(220)_{pc}$ peak disappears but the best fit is observed for the two phase model only, as can be seen from the $\chi^2$ values for the Cm, P4mm and Cm+P4mm structural models given in Fig. 8(c), (d) and (e), respectively, for a representative temperature of 450K. This is in agreement with the FWHM analysis presented in Fig.6 also. To confirm the structure(s) for temperatures in between 250-525K, over which the triplet like feature of $(220)_{pc}$ disappears but additional broadening persists (see Fig. 6), we first considered single phase Cm and single phase P4mm space groups for a representative temperature of 450K and the fits are shown in Fig. 8(c) and (d). Both the single phase models are not able to account for the $(002)_{pc}$ tetragonal peak position. We therefore considered the phase coexistence model (P4mm+Cm) for this temperature. This improves the quality of fits and reduces the $\chi^2$ value significantly. The two phase model was found necessary up to 525K after which the pseudotetragonal monoclinic phase disappears leaving behind the pure tetragonal phase as confirmed by the Rietveld refinements. Figs. 8(f) and (g) show the Rietveld fits using P4mm space group and $Pm\bar{3}m$ space group at 550K and 800K, respectively.

Figs. 9 and 10 depict the temperature evolution of the $(111)_{pc}$, $(200)_{pc}$ and $(220)_{pc}$ peaks of PSZT for x=0.530 (PSZT530) and 0.535 (PSZT535), respectively. As shown elsewhere[6] for PSZT530, refinements using SXRD data have shown that the pseudotetragonal monoclinic phase coexists with a pseudorhombohedral monoclinic phase[6], both in the Cm space group, at room temperature and below until up to ~260K. At T≲260K, there is a Cm to Cc antiferrodistortive phase transition, the signature of which is seen in the neutron diffraction patterns only through the appearance of new superlattice peaks[6,7] and not in the SXRD patterns. The SXRD patterns of PSZT530 below 260K are therefore not different from these above this temperature upto about 400K. For T>400K, the $(220)_{pc}$ profile looks like a doublet as expected for the tetragonal phase but the additional broadening of the $(111)_{pc}$ rules this out. The temperature dependence of the FWHM of the $(111)_{pc}$ and $(200)_{pc}$ peaks are shown in Fig. 11 and 12 for PSZT530 and PSZT533, respectively. It is evident from these figures that for temperatures greater than 600K, FWHM of the $(200)_{pc}$ peak is nearly constant. Similarly, the FWHM of $(111)_{pc}$ peak is constant for T$\gtrsim$550K. Both these features imply that the tetragonal structure of PSZT530 and PSZT535 is stable in the temperature range 550K$\lesssim$T$\lesssim$625K. Below 550K, the broadening of the $(111)_{pc}$ peak starts growing suggesting the appearance of the monoclinic phases in Cm space group as a coexisting phase. Lebail fits for PSZT530 have been shown in our previous work (Ref. 6) which confirmed that the pseudotetragonal and pseudorhombohedral monoclinic phases coexist in PSZT530 from low temperatures upto T$\lesssim$550K and transform to pure tetragonal phase at T$\simeq$550K. Thus the sequence of phase transitions for PSZT530 and PSZT535 is: pseudotetragonal monoclinic (Cm) + pseudorhombohedral monoclinic (Cm) to tetragonal (P4mm) and then tetragonal to cubic (Pm$\bar{3}$m).

Evolution of the $(111)_{pc}$, $(200)_{pc}$ and $(220)_{pc}$ profiles with temperature for PSZT with x=0.545 (i.e., PSZT545) is shown in Fig. 13. The splitting of both $(111)_{pc}$ and $(200)_{pc}$ and anomalous broadening of $(200)_{pc}$ profiles are consistent with the pseudorhombohedral monoclinic phase in Cm space group.[3] The splitting of the $(111)_{pc}$ and $(220)_{pc}$ peaks disappears around 475K but the FWHM of the $(111)_{pc}$ and $(200)_{pc}$ peaks persists and decreases with increasing temperature as shown in Fig.14 (a) and (b). The presence of additional broadening in the $(200)_{pc}$ peak and the disappearance of the broadening of the $(111)_{pc}$ peak in the temperature range 550K$\leq$T$\leq$600K suggests that $(200)_{pc}$ is still a doublet while $(111)_{pc}$ is a singlet in this temperature range. Both these imply tetragonal structure in the temperature range 550K$\lesssim$T$\lesssim$600K. Above 600K, the broadening of the $(200)_{pc}$ peak also disappears implying singlet nature of this and other peaks. This suggests that the structure is cubic for T>600K. The broadening of the $(111)_{pc}$ peak below T≃550K eventually leading to splitting of the peak at lower temperatures confirms that the pseudorhombohedral monoclinic (Cm) phase of PSZT545 transforms to the tetragonal (P4mm) phase at 525K≲T≲550K and this tetragonal phase at still higher temperatures (T≳600K) transforms to the cubic ($Pm\bar{3}m$) phase.

Rietveld refinements for PSZT with x=0.550 (PSZT550)[7] have confirmed pure Cm phase structure at room temperature and it transforms to the Cc phase below room temperature through an AFD transition confirmed by neutron diffraction measurements[7]. The temperature dependence of $(111)_{pc}$, $(200)_{pc}$ and $(220)_{pc}$ peak profiles of PSZT550 is shown in Fig. 15. It is evident from this figure that the splitting of the peaks disappears above 450K. However, the width of the peaks is still quite large and decreases with increasing temperature as shown in Fig. 16 for the $(111)_{pc}$ and $(200)_{pc}$ peaks. This suggests that none of the peaks is a singlet. However, at T>600K, the FWHM of the both peaks is nearly temperature independent suggesting that all the peaks are now singlet as expected for the cubic phase. These results show that the pseudorhombohedral monoclinic (Cm) phase of PSZT550 transforms directly into the cubic phase.

It is important to reiterate that all the PSZT compositions discussed above undergo an antiferrodistortive (AFD) phase transition below room temperature leading to appearance of superlattice peak in neutron powder diffraction patterns only. As a result of this AFD transition, the Cm space group structure changes to Cc space group structure [6,7]. Since these superlattice peaks are not discernible in SXRD patterns below room temperature[6,7], the SXRD data does not reveal the Cm to Cc AFD transition below room temperature. We have, therefore, in the discussions above, not distinguished between Cc and Cm space groups even at 100K. Further, the ferroelectric phase transition temperatures decrease slightly with increasing $Zr^{4+}$ content in the composition range $0.515 \leq x \leq 0.550$ from about 614K for x=0.515 to 604K for 0.550. However, since SXRD data were collected at 25K interval, several compositions appear to show transitions nearly at the same temperature as per the SXRD data.

From the above discussion, it is evident that the high temperature cubic phase first transforms into a tetragonal phase for the PSZT composition with $x \leq 0.545$ whereas it transforms directly into a pseudorhombohedral monoclinic phase for x=0.550. This suggests the existence of a triple point at $0.545 \lesssim x \lesssim 0.550$. The variation of unit cell parameters and unit cell volume over a limited temperature range, as obtained by Rietveld refinement of the structure at various temperatures, is shown in Fig.17 for the cubic to tetragonal phase transition of PSZT with (a) x=0.515, (b) x=0.520, (c) x=0.525, (d) x=0.530, (e) x=0.535 and (f) x=0.545. It is evident from these figures that the cubic to tetragonal phase transition is accompanied with a discontinuous change in the unit cell volume at the transition temperature. This implies first order nature of the tetragonal to cubic phase transition in agreement with the observation of thermal hysteresis in the dielectric measurements discussed earlier. The vertical lines at $T_c$ are drawn to show the value of discontinuous change in the unit cell volume ($\Delta V$) at $T_c$ for tetragonal to cubic phase transition determined by dielectric measurements during heating. Fig. 2(b) shows the discontinuous change ($\Delta V$) in the unit cell volume at the cubic to tetragonal phase transition temperature as a function of composition. It is evident from this figure that the extrapolated value of $\Delta V$ approaches zero for $x \simeq 0.550$. This is in agreement with the continuous variation of the unit cell volume across the cubic to monoclinic phase transition shown in Fig. 17(g) for x=0.550. The vanishing of ΔV at x≃0.550 coincides with the vanishing of the thermal hysteresis ΔT= ($T_c^H$-$T_c^C$) at x≃0.553 in the dielectric studies (Fig. 2(a)). Both these clearly reveal a crossover from first order to second order phase transition with a tricritical point at x≃0.550, which is also the triple point in the phase diagram of PSZT.

**C. Landau-Devonshire theory considerations**

We have determined the Landau-Devonshire free energy curves as a function of composition with a view to visualize the flattening of the free energy surface associated with the cubic to tetragonal phase transition on approaching the MPB from the tetragonal side. For this, we have considered Landau-Devonshire free energy function used by Haun et al[22] to calculate the free energy of tetragonal phase of PZT. For the FE tetragonal phase with polarization components $P_1$=$P_2$=0, $P_3$≠0, free-energy function can be written as follows, under zero stress condition, [22]:

$$\Delta G = \alpha_1 P_3^2 + \alpha_{11} P_3^4 + \alpha_{111} P_3^6 \quad (1)$$

The coefficients $\alpha_{11}$ and $\alpha_{111}$ are assumed to be temperature independent. $\alpha_{11}$ is positive for second-order phase transition and negative for first-order phase transition, while $\alpha_{111}$ is always positive. The coefficient $\alpha_1$ is assumed to have the following type of temperature dependence based on the Curie-Weiss law:

$$\alpha_1 = (T-T_0)/(2\varepsilon_0 C) = \alpha_0(T-T_0) \quad (2)$$

where $\varepsilon_0$, C and $T_0$ are permittivity of free space, Curie constant and Curie-Weiss temperature, respectively. Now using the minimization condition $\partial G/\partial P_3 = 0$ and ΔG=0 at $T_c$, one obtains the following expressions for the remaining two coefficients[22]:

$$\alpha_{11} = -3(T_c - T_0)/(\varepsilon_0 C P_{3c}^2) \quad (3)$$

and

$$\alpha_{111} = -3(T_c - T_0)/(2\varepsilon_0 C P_{3c}^4) \quad (4)$$

where $P_{3c}$ is the value of spontaneous polarization at $T_c$. For a first-order phase transition $T_c$ is related to $T_0$ in the following manner:

$$T_c = T_0 + \alpha_{11}^2/(12\alpha_0\alpha_{111}) \quad (5)$$

Minimization condition $\partial G/\partial P_3 = 0$ yields the following expression for spontaneous polarization

$$P_3^2 = [-\alpha_{11} + \sqrt{\alpha_{11} - 3\alpha_1\alpha_{111}}]/3\alpha_{111} \quad (6)$$

Substituting values of $\alpha_1$, $\alpha_{11}$ and $\alpha_{111}$ in above relation, we obtain the following expression for the spontaneous polarization[22]:

$$P_3^2 = \psi P_{3C}^2 \text{ where } \psi = 2/3\{1 + [1 - \frac{3(T-T_0)}{4(T_0-T_c)}]^{1/2}\} \quad (7)$$

The spontaneous transformation strains ($x_i = \partial G/\partial X_i$) under zero stress condition were obtained using the procedure given by Haun et al[22]:

$$x_1 = Q_{12} P_3^2 \quad (8)$$

$$x_3 = Q_{11} P_3^2 \quad (9)$$

Combining the above with eq. (7), we get:

$$x_1 = x_{1c}\psi \text{ where } x_{1c} = Q_{12} P_{3c}^2 \text{ and} \quad (10)$$

$$x_3 = x_{3c}\psi \text{ where } x_{3c} = Q_{11} P_{3c}^2 , \quad (11)$$

where $x_{1c}$ and $x_{3c}$ are the values of $x_1$ and $x_3$ at $T_c$.

Thus ΔG as a function of temperature can be calculated from the knowledge of $x_3$, $T_c$, $T_0$ and C, all of which can be determined from the experimental data (see Haun et al[22]), and $Q_{11}$. We now illustrate the determination of the coefficients in Eq. (1) using

PSZT515 as an example. Lattice parameters corresponding to the cubic to tetragonal phase transition for PSZT515 are shown in Fig. 18(a). Lattice parameter $a_c$ of the high temperature cubic phase was extrapolated into the tetragonal region to obtain the extrapolated value $a_c'$ which was used to calculate the spontaneous strain $x_1$ and $x_3$ for the tetragonal phase. The extrapolated values are shown with open circles in the Fig. 18(a). The symmetry adopted spontaneous strains associated with the cubic to tetragonal phase transition are given by

$$x_1 = \frac{a_T - a_c'}{a_c'} \quad \text{and} \tag{12}$$

$$x_3 = \frac{c_T - a_c'}{a_c'} \tag{13}$$

These were determined using the data given in Fig. 18(a) and are plotted in Fig. 18(b). From Eqs. 10 and 11, it is evident that the spontaneous strain is related to the temperature dependent function $\psi$. The dependence of $\psi$ on temperature is given by the following relationship[22]:

$$\psi = \frac{2}{3}\{1 + [1 - \frac{3(T - T_0)}{4(T_0 - T_c)}]^{1/2}\}. \tag{14}$$

$\psi$ is thus a function of $T_c$ and $T_0$ where $T_c$ is the phase transition temperature and $T_0$ is the Curie-Weiss temperature. To calculate $T_0$, we have plotted $\frac{1}{\varepsilon'}$ vs T in Fig. 18(c) which also depicts Curie-Weiss fit to the measured $\varepsilon'(T)$ data for PSZT515 above $T_c$. Since $x_3$ and $\psi$ are related by $x_3 = x_{3c}\psi$, by fitting the spontaneous strain data ($x_3$) with $\psi$, we can calculate $x_{3c}$. Further, $x_{3c} = Q_{11} P_{3c}^2$ and $P_3^2 = \psi P_{3c}^2$ (see Eq. 7 and 11). For $Q_{11}$, we have used the values given for PZT from ref. 31 and calculated $P_{3c}^2$. The variation of spontaneous polarization ($P_3^2$), so obtained, with temperature is plotted in Figs. 18(d). For the compositions with x=0.520 and 0.525, similar procedure was adopted to calculate the various parameters. Figs. 19 and 20 show the variation of lattice parameters, spontaneous strain, Curie-Weiss fit and spontaneous polarization as a function of temperature corresponding to the tetragonal to cubic phase transition for PSZT520 and 525, respectively.

We have calculated Landau coefficients using Eq. 2, 3 and 4. Various parameters used in Landau theory calculations are given in Table I. Using the Landau coefficients so determined, we calculated the free energy density. Free energy density as a function of polarization at several temperatures for PSZT515, 520 and 525 are plotted in *Fig. 21*. The energy barrier ($\Delta G_{Tc}$) corresponding to cubic to tetragonal thermodynamic phase transition temperature, at which one observes triply degenerate minima in ΔG, were obtained from the free energy plots. For a tricritical transition this energy barrier $(\Delta G)_{Tc}$ should vanish making the free energy surface flat at the phase transition temperature. The variation of $(\Delta G)_{Tc}$ with composition is shown in Fig. 2(c). On extrapolating the linear behavior in Fig. 2(c), we find that the energy barrier $(\Delta G)_{Tc}$ vanishes for x≃0.550. This reveals the flattening of the energy surface at $T_c$ for x≃0.550, as expected for a tricritical point predicted on the basis of dielectric and SXRD studies also. Thus Landau theory calculations also confirm the findings based on vanishing of thermal hysteresis in dielectric studies and the discontinuous change in unit cell volume in SXRD studies that there exists a tricritical point at x≃0.550 which is also the triple point in the PSZT phase diagram.

As evident from eq. (8) and (9), $Q_{11}$ and $Q_{12}$ can also be obtained from the temperature dependence of spontaneous polarization ($P_3$) and spontaneous strain. We therefore calculated the spontaneous polarization using the atomic displacements $\xi_i$ and Born effective charges $Z_i$ for the cubic phase using the following relationship:

$$P = \frac{e}{V}\sum_i Z_i\, \xi_i \qquad (15)$$

where e is the charge of electron, V is the volume of the unit cell and $\xi_i$ the positional coordinates of the cations and anions obtained by Rietveld refinements. For the tetragonal phase of PSZT515 in P4mm space group, Pb/Sr atoms were kept fixed at (0, 0, 0) position, they therefore do not contribute to the polarization. Values for Born effective charges for Zr, Ti and oxygen were taken from Ref. 32. Polarization for PSZT515 has been calculated from the Rietveld refined coordinates and Born effective charges and has been plotted as a function of temperature in Fig. 22(a). In our calculations, for polarizations and spontaneous strains, we have used data from room temperature to high temperatures because of a phase transition to the monoclinic phase (see section D. phase diagram of PSZT and Fig. 23) just below the room temperature clear signature of two transitions below room temperature have been observed in the attenuation of sound velocity[33] but the diffraction pattern do not reveal such a transition. Similar situation exists in PZT515 also[34]. Variation of spontaneous strains ($x_1$ and $x_3$) for PSZT515 with temperature calculated from eqns. 12 and 13 and temperature dependence of lattice parameters has been already shown in Fig. 18(b). Fig. 22(b) depicts the variation of spontaneous strains ($x_1$ and $x_3$) with $P_3^2$. It shows linear dependence. We therefore use the relation $x_1=Q_{12}\,P_3^2$ and $x_3=Q_{11}\,P_3^2$ to fit the variation depicted in Fig. 22(b). Unlike thin films, where one has to consider higher order electrostrictive coupling terms to fit the dependence of spontaneous strain on spontaneous polarization[35], our results show a nice linear behavior in Fig. 22(b). *Values of $Q_{11}$ and $Q_{12}$ so obtained using straight line fits are*

*$4.2*10^{-2}$ $m^4/C^2$ and $-1.56*10^{-2}$ $m^4/C^2$ which are less than that of pure PZT515 ($Q_{11}$ for pure PZT = $9.468*10^{-2}$ $m^4/C^2$ and $Q_{12}$= $-4.468*10^{-2}$ $m^4/C^2$)* but are of the same order of magnitude. This difference in the $Q_{11}$ and $Q_{12}$ values may be due to 6% Sr doping in PZT or due to the inaccurate polarization values since X-rays are not ideally suited to locate accurately the lighter atoms like oxygen. We have calculated the values of $P_{3c}^2$ using $Q_{11}$ and $Q_{12}$, which are found to be $21.68*10^{-2}$ and $21.67*10^{-2}$ $C^2/m^4$, respectively. As expected both the values of $P_{3c}^2$ are nearly equivalent. It is therefore evident that one can use $Q_{11}$ or $Q_{12}$ to calculate the free energy. In our case we have chosen $Q_{11}$ to calculate the free energy.

For the two phase compositions, such as PSZT520 and PSZT525 which transform to the pure tetragonal phase for temperatures above 525, the stability region of the pure tetragonal phase is rather narrow (550K≤T≤600K) and since we collected SXRD data at 25K interval, we did not have sufficient data points to carry out the type of analysis presented in Fig.22. Further, the temperature variation of $P_3^2$ and $x_3$ over such a narrow temperature range close to $T_c$ may not reflect the ideal functional form and may therefore introduce errors in the Q values. Since the change in Q with small variation in x from 0.515 to 0.520 and to 0.525 is rather small for PZT, we assume the Q value of PSZT515 for PSZT520 and PSZT525 also and carried out the free energy calculations. Fig. 2 (d) depicts the variation of $(\Delta G)_{Tc}$ with x. It is evident from the figure that the dependence shown in Fig. 2(c) and (d) are qualitatively similar and reveal flatterning of the energy surface with increasing x. The extrapolated $x_c$ value at which $(\Delta G)_{Tc}$ vanishes comes out to be $x \simeq 0.546$ using the new set of values of electrostrictive coefficients which is very close to $x_c \simeq 0.550$ obtained using electrostrictive coefficients of pure PZT. Thus our assumption to use $Q_{11}$ values of pure PZT for PSZT also is justifiable. We further compare the Landau coefficients of PSZT obtained using above two approaches to that of pure PZT in Table II. The Landau coefficients are of the same order of magnitude as that of pure PZT using above two approaches. Based on the results shown in Fig. 2(c) and (d), we may conclude that a tricritical point exists in PSZT for $0.545<x\lesssim 0.550$ and it coincides with the triple point in the PSZT phase diagram.

**D. Phase diagram of PSZT around MPB region:**

In pure PZT, MPB occurs at $x\approx 0.520$ for which the piezoelectric and dielectric responses are maximum[1,18,3] whereas it occurs at x=0.530 for PSZT[6(a)]. On the basis of the present high resolution SXRD and dielectric studies in conjunction with earlier neutron powder diffraction and sound velocity studies[6(a),7], we have constructed a temperature (T) composition (x) phase diagram of PSZT around the MPB region shown in Fig. 23. It is evident from this figure that the ferroelectric to paraelectric phase transition temperature decreases with decreasing $PbZrO_3$ content, as expected for the lower transition temperature for $PbZrO_3$. The low temperature phase boundary corresponds to the antiferrodistortive (AFD) phase transition and has been drawn using sound velocity measurements which reveal anomalies at the phase transition temperature[33]. The AFD phase transition temperature increases on increasing the Zr-content. This shows the weakening of driving force for AFD transition with decreasing $Zr^{4+}$ content on approaching the MPB. For x=0.515, using Rietveld refinement of SXRD data we found that structure is tetragonal in the temperature range $100K\leq T\leq 614K$ and for $T>614K$ the structure is cubic. But sound velocity measurements confirm that this composition also undergoes a low temperature phase transition around ~180K. In the absence of any additional splitting of the $(220)_{pc}$ peak or splitting of $(111)_{pc}$ peak below 180K (see Fig. 3), it is not possible to guess the structure of the low temperature phase of PSZT515. A similar situation has been pointed out by Ragini[34] for pure PZT with x=0.515. The space group of the low temperature phase of PSZT515 may be I4cm[36] or Cc[2] but more work is needed for making a choice. For 0.520≲x≲0.525 tetragonal and pseudotetragonal (PT) phases in the P4mm and Cm space groups coexist, while for 0.525<x≲0.535 pseudotetragonal (PT) and pseudorhombohedral (PR) monoclinic phases both in Cm space group represent the phase coexistence region. For x=0.545, the pseudorhombohedral monoclinic phase in Cm space group is stable up to 525K above which it transforms to the tetragonal phase and finally to the cubic phase. For x=0.550 pseudorhombohedral monoclinic phase transforms directly to the cubic phase. The triple point where the paraelectric to ferroelectric boundary and MPB intersect occurs at 0.545<x≲0.550. The tricritical point in PSZT also occurs at 0.545<x≲0.550. The earlier work of Mishra et al[18] revealed a tricritical point at x=0.550 for pure PZT also. This shows that the MPB in PSZT is less tilted, i.e. more vertical than pure PZT. Vertical MPB implies that temperature variation will always keep the material close to the MPB. The vertical MPB[37] is ideally preferred over tilted as the flat energy surface associated with the tricritical point may be nearly preserved even at room temperature. The more vertical nature of the MPB in PSZT may be the reason for higher value of $d_{33}$~300pC/N in PSZT in comparison to pure PZT for which it is ~180pC/N[38].

## IV. CONCLUSIONS

We have investigated the nature of ferroelectric phase transition using temperature dependent dielectric and SXRD studies for compositions around the MPB of PSZT. Using dielectric studies, we have shown that Curie transition is of first order for $x \leq 0.545$, whereas it has a second order character for $x \gtrsim 0.550$. Using Rietveld analysis of temperature dependent SXRD data, we have shown that the discontinuous change ($\Delta V$) in the unit cell volume at the cubic to tetragonal phase transition decreases with increasing x and finally vanishes for $x \lesssim 0.550$, revealing a crossover from first order character of the cubic transition for $x \lesssim 0.545$ to second order at $x \simeq 0.550$. The dielectric and SXRD results confirm the existence of a tricritical point for $0.545 < x \lesssim 0.550$, at which first order cubic to tetragonal line of phase transition, second order cubic to psuedorhombohedral monoclinic line of phase transition and MPB intersect and give rise to a triple point. Free energy barrier $(\Delta G)_{Tc}$ between the coexisting tetragonal and cubic phases at the thermodynamic phase transition temperature calculated using Landau-Devonshire theory for cubic to tetragonal phase transition of PSZT, decreases linearly and vanishes around $x \simeq 0.550$ as expected for a tricritical point. These calculations confirm the flattening of free energy surface around $x \simeq 0.550$. The flattening of free energy surface at high temperatures leads to the flatter energy surface at room temperature also[11] for MPB compositions and results in enhanced electromechanical[39] and piezoelectric response[38] in the vicinity of MPB for PSZT[38] as expected on the basis of the polarization rotation mechanism[8, 11].

**ACKNOWLEDGMENTS**

DP and YK acknowledge financial support from Department of Science and Technology (DST), Govt. of India and Japan Society for the Promotion of Science (JSPS) of Japan under the Indo-Japan Science Collaboration Program. The synchrotron radiation experiments were performed at the BL02B2 beamline of Spring-8 with the approval of Japan Synchrotron Radiation Research Institute (Proposal Nos. 2011A1324 and 2011A0084). DP acknowledges financial support from DST, Government of India, for J.C. Bose National Fellowship grant.

## Figure Captions:

**Fig.1.** (Color online) Temperature dependence of dielectric permittivity (ε′) of $(Pb_{0.94}Sr_{0.06})(Zr_xTi_{1-x})O_3$ at 100 kHz for the compositions with (a) x = 0.515, (b) x = 0.520, (c) x = 0.525, (d) x = 0.530, (e) x=0.535, (f) x = 0.545 and (g) x = 0.550 during heating (filled circles) and cooling (open circles) cycles.

**Fig.2.** (Color online) (a) The cubic to tetragonal phase transition temperature ($T_c$) during heating and cooling cycles, (b) discontinuous change (ΔV) in the unit cell volume at the transition temperature $T_c$ during heating, (c) Landau free energy barrier at $T_c$ between the cubic and tetragonal phases of PSZT for x=0.515, 0.520 and 0.525 calculated using Landau-Devonshire theory for which we used electrostrictive coupling coefficient ($Q_{11}$) of PZT and (d) Landau free energy barrier at $T_c$ between the cubic and tetragonal phases of PSZT for x=0.515, 0.520 and 0.525 calculated using Landau-Devonshire theory for which the value of electrostrictive coupling coefficient $Q_{11}$ was obtained from the structural parameters of PSZT515.

**Fig.3.** (Color online) The evolution of synchrotron powder XRD profiles of the $(111)_{pc}$, $(200)_{pc}$ and $(220)_{pc}$ reflections of tetragonal PSZT515 with temperature.

**Fig.4.** (Color online) The evolution of synchrotron powder XRD profiles of the $(111)_{pc}$, $(200)_{pc}$ and $(220)_{pc}$ reflections of PSZT520 with temperature.

**Fig.5.** (Color online) The evolution of synchrotron powder XRD profiles of the $(111)_{pc}$, $(200)_{pc}$ and $(220)_{pc}$ reflections of PSZT525 with temperature.

**Fig.6.** (Color online) The variation of FWHM of the (a) $(111)_{pc}$ and (b)$(200)_{pc}$ reflections of PSZT520 with temperature.

**Fig.7.** (Color online) The variation of FWHM of the (a) $(111)_{pc}$ and (b) $(200)_{pc}$ reflections of PSZT525 with temperature.

**Fig.8.** (Color online) Observed (dots), calculated (continuous line), and difference (bottom line) profiles of selected $(111)_{pc}$, $(200)_{pc}$, and $(220)_{pc}$ reflections for PSZT520 at various temperatures obtained after Rietveld refinement using plausible structural models. The vertical tick marks above the difference profiles give the positions of the Bragg reflections.

**Fig.9.** (Color online) The evolution of synchrotron powder XRD profiles of the $(111)_{pc}$, $(200)_{pc}$ and $(220)_{pc}$ reflections of PSZT530 with temperature.

**Fig.10.** (Color online) The evolution of synchrotron powder XRD profiles of the $(111)_{pc}$, $(200)_{pc}$ and $(220)_{pc}$ reflections of PSZT535 with temperature.

**Fig.11.** (Color online) The variation of FWHM of the (a) $(111)_{pc}$ and (b) $(200)_{pc}$ reflections of PSZT530 with temperature.

**Fig.12.** (Color online) The variation of FWHM of the (a) $(111)_{pc}$ and (b) $(200)_{pc}$ reflections of PSZT535 with temperature.

**Fig.13.** (Color online) The evolution of synchrotron powder XRD profiles of the $(111)_{pc}$, $(200)_{pc}$ and $(220)_{pc}$ reflections of PSZT545 with temperature.

**Fig.14.** (Color online) The variation of FWHM of the (a) $(111)_{pc}$ and (b) $(200)_{pc}$ reflections of PSZT545 with temperature.

**Fig.15.** (Color online) The evolution of synchrotron powder XRD profiles of the $(111)_{pc}$, $(200)_{pc}$ and $(220)_{pc}$ reflections of PSZT550 with temperature.

**Fig.16.** (Color online) The variation of FWHM of the (a) $(111)_{pc}$ and (b) $(200)_{pc}$ reflections of PSZT550 with temperature.

**Fig.17.** (Color online) Variation of (a) a and c lattice parameters and (b) unit cell volume of various PSZT compositions with temperature. Error bars are less than the size of the data points. *Continues lines are guide to eyes.*

**Fig.18.** (Color online) Temperature dependence of (a) the a and c lattice parameters (open dots represent the extrapolated values of $a_c$ in the tetragonal region *and continues lines are guide to eyes)*, (b) spontaneous strains $x_1$ and $x_3$ *(continues lines represent least squares fit to the data points using a second order polynomial)*, (c) dielectric stiffness ($\frac{1}{\varepsilon'}$) (the continuous line gives the Curie-Weiss fit) and (d) $P_3^2$ for PSZT515 *(continues lines are guide to eyes)*.

**Fig.19.** (Color online) (a) Temperature dependence of (a) the a and c lattice parameters (open dots represent the extrapolated values of $a_c$ in the tetragonal region *and continues lines are guide to eyes)*, (b) spontaneous strains $x_1$ and $x_3$ *(continues lines represent least squares fit to the data points using a second order polynomial)*, (c) dielectric stiffness ($\frac{1}{\varepsilon'}$) (the continuous line gives the Curie-Weiss fit) and (d) $P_3^2$ for PSZT520 *(continues lines are guide to eyes)*.

**Fig.20.** (Color online) (a) Temperature dependence of (a) the a and c lattice parameters (open dots represent the extrapolated values of $a_c$ in the tetragonal region and *continues lines are guide to eyes)*, (b) spontaneous strains $x_1$ and $x_3$ *(continues lines represent least squares fit to the data points using a second order polynomial)*, (c) dielectric stiffness ($\frac{1}{\varepsilon'}$) (the continuous line gives the Curie-Weiss fit) and (d) $P_3^2$ for PSZT525 *(continues lines are guide to eyes)*.

**Fig.21.** (Color online) Gibbs free energy profiles of PSZT515 calculated at different temperatures using Landau-Devonshire theory (Ref.22). $T_2$ is the characteristic temperature at which the inflexion point in the free energy profile develops while for the temperature $T_1$, a local minimum for the tetragonal phase appears well above the thermodynamic phase transition temperature $T_c$ at which the free energy profile is triply degenerate on account of coexistence of the tetragonal and cubic phases. At temperature well below $T_c$, the minima corresponding to the tetragonal phase become global minima.

***Fig.22.*** *(Color online) (a) Temperature dependence of $P_3^2$ and (b) variation of spontaneous strains ($x_1$, $x_3$) with $P_3^2$ for PSZT515. T*he continuous line gives the fit using the relation given in eqns. 8 and 9. *$P_3^2$ has been calculated from Born effective charges and atomic coordinates using Eqn. 15.*

**Fig.23.** (Color online) Schematic phase diagram of PSZT around the MPB region constructed on the basis of present work and earlier neutron and ultrasonic measurements [Ref. 6 and 7] on the same set of PSZT samples. Open circles in the phase diagram correspond to the high temperature cubic ($Pm\bar{3}m$) to tetragonal (P4mm) phase transition temperature during heating, obtained from dielectric measurements. The filled circles correspond to low temperature antiferrodistortive phase (space group Cc) transition temperatures [Ref. 6 and 7]. The symbols PT and PR represent pseudotetragonal and pseudorhombohedral monoclinic phases, respectively. The vertical lines are first order phase boundaries across which the two neighbouring phases coexist. Also, the inclined P4mm-Cm (PT)/Cm (PR) phase boundary is a first order phase boundary. The phase boundary marked with filled circle is linked with the AFD transition whose order (I or II) is unsettled.

## Table Captions:

**Table I:** Parameters obtained using the temperature dependence of cell parameters and dielectric constant. Last column shows the energy barrier ($\Delta G$) between the cubic and tetragonal phases at the transition temperature ($T_c$).

**Table II:** Landau coefficients of (i) PZT, (ii) PSZT obtained using $Q_{11}$ values of PZT and (iii) PSZT obtained using $Q_{11}$ value of PSZT515.

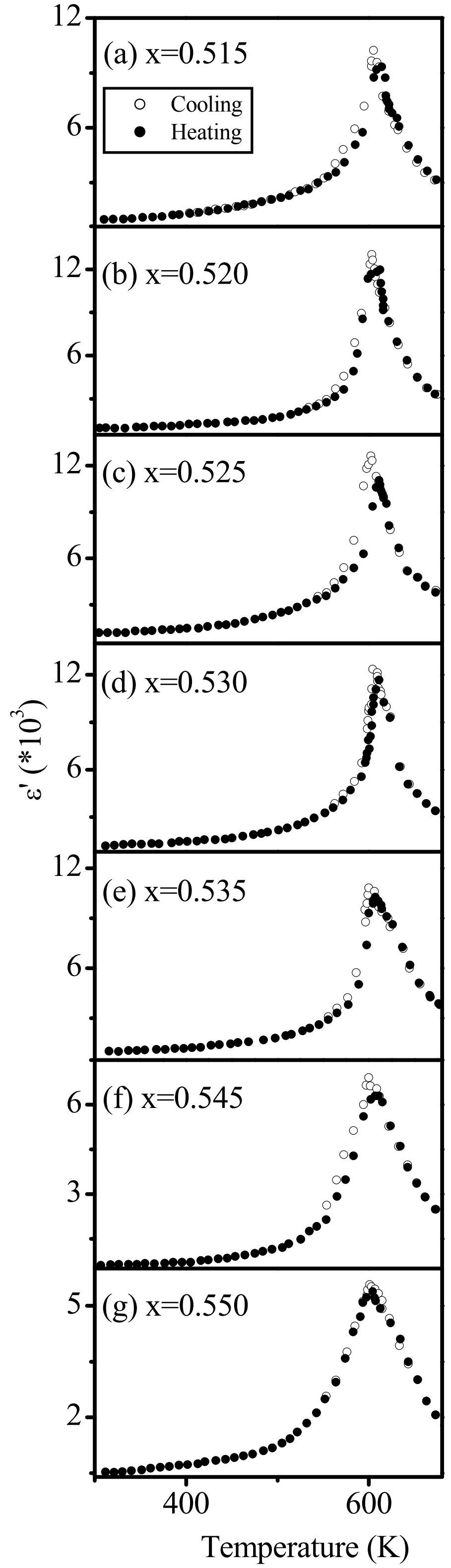

**Fig.1**

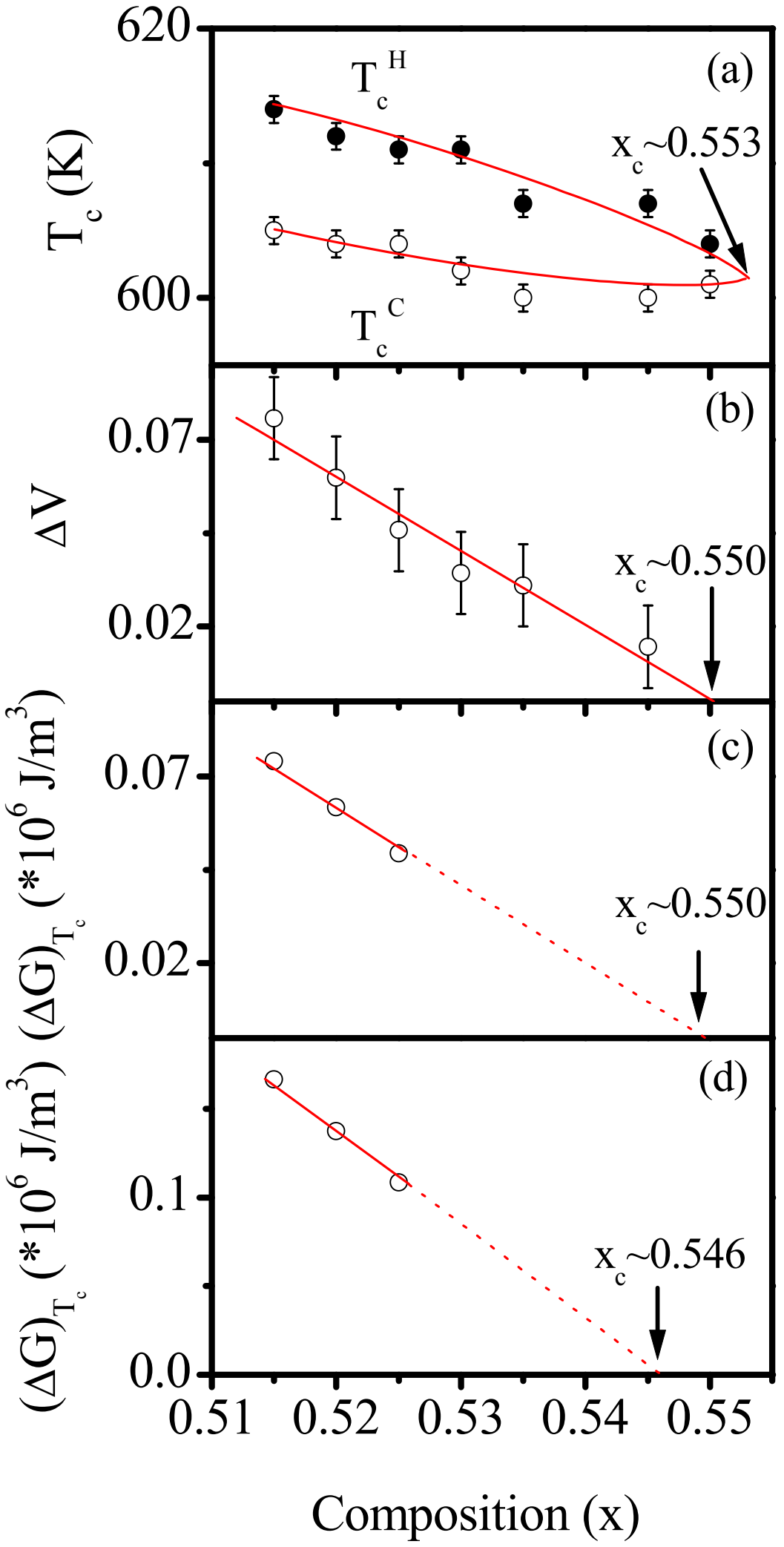

**Fig.2**

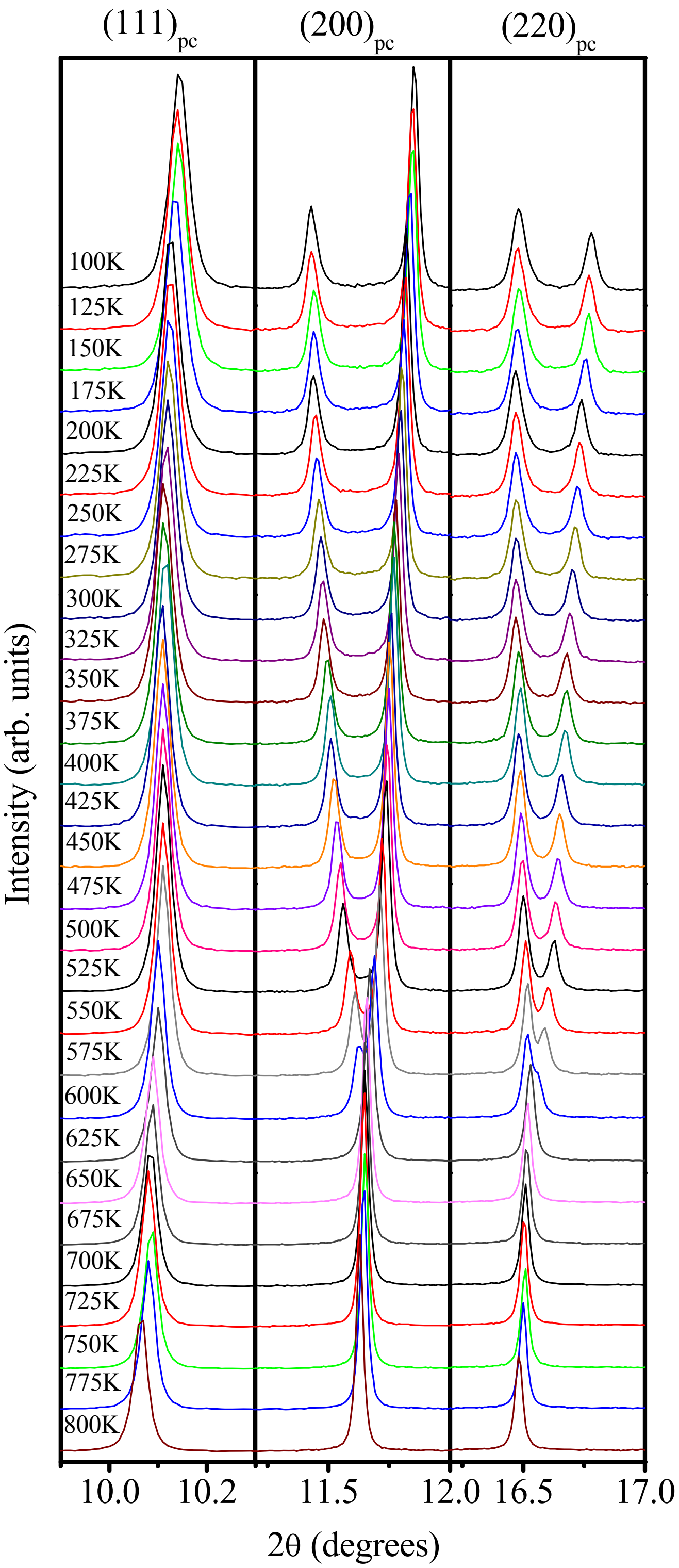

**Fig.3**

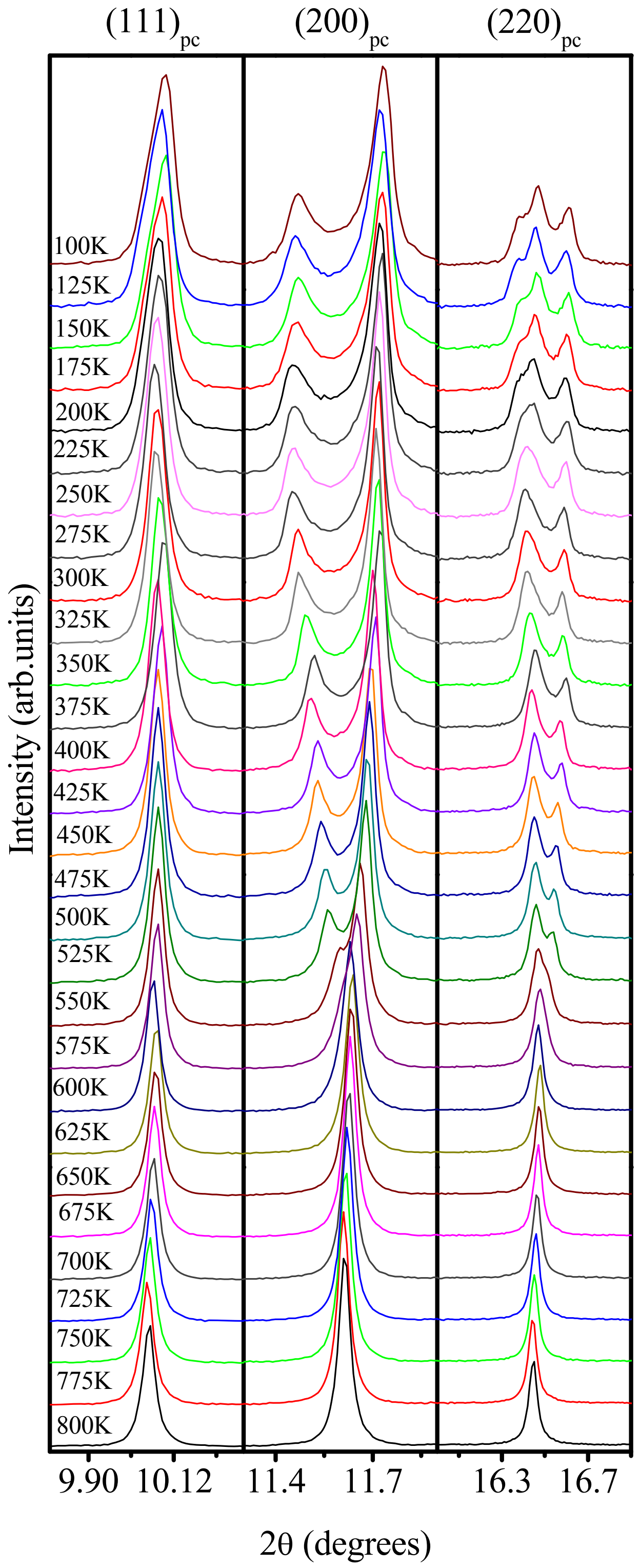

**Fig.4**

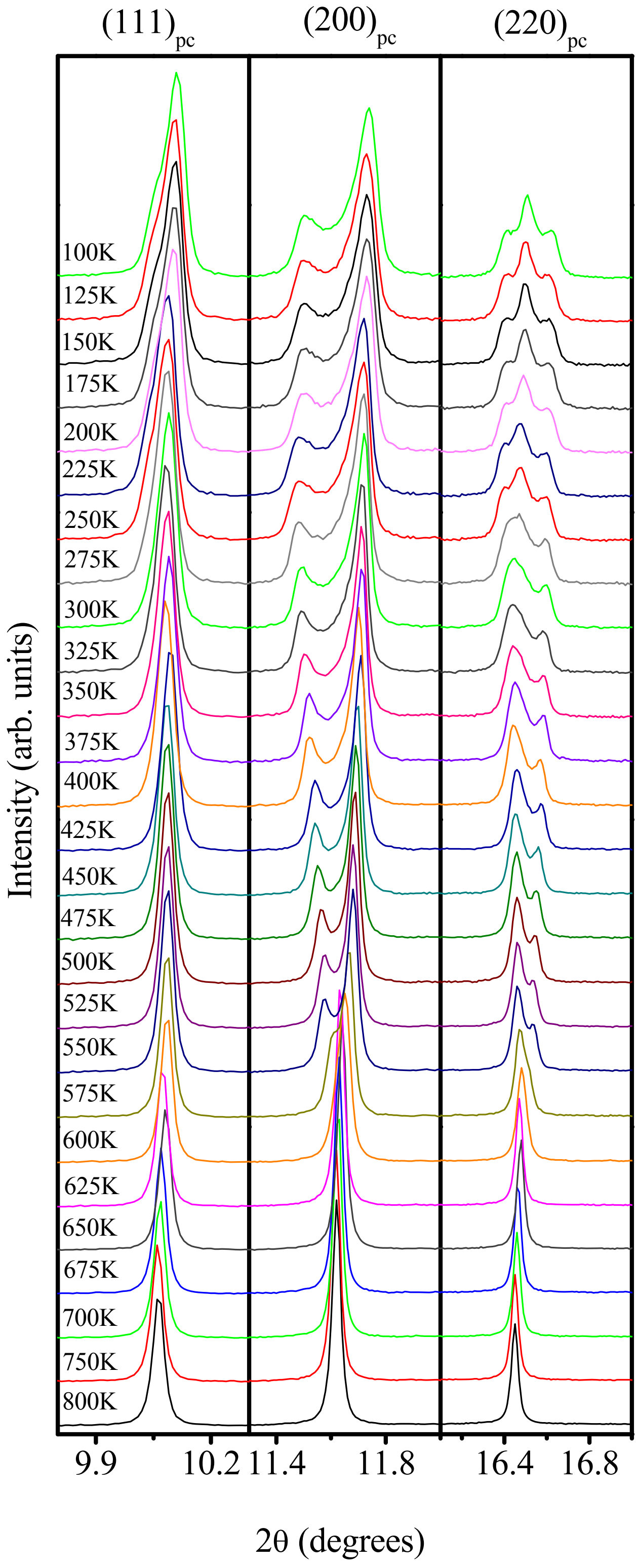

***Fig.5***

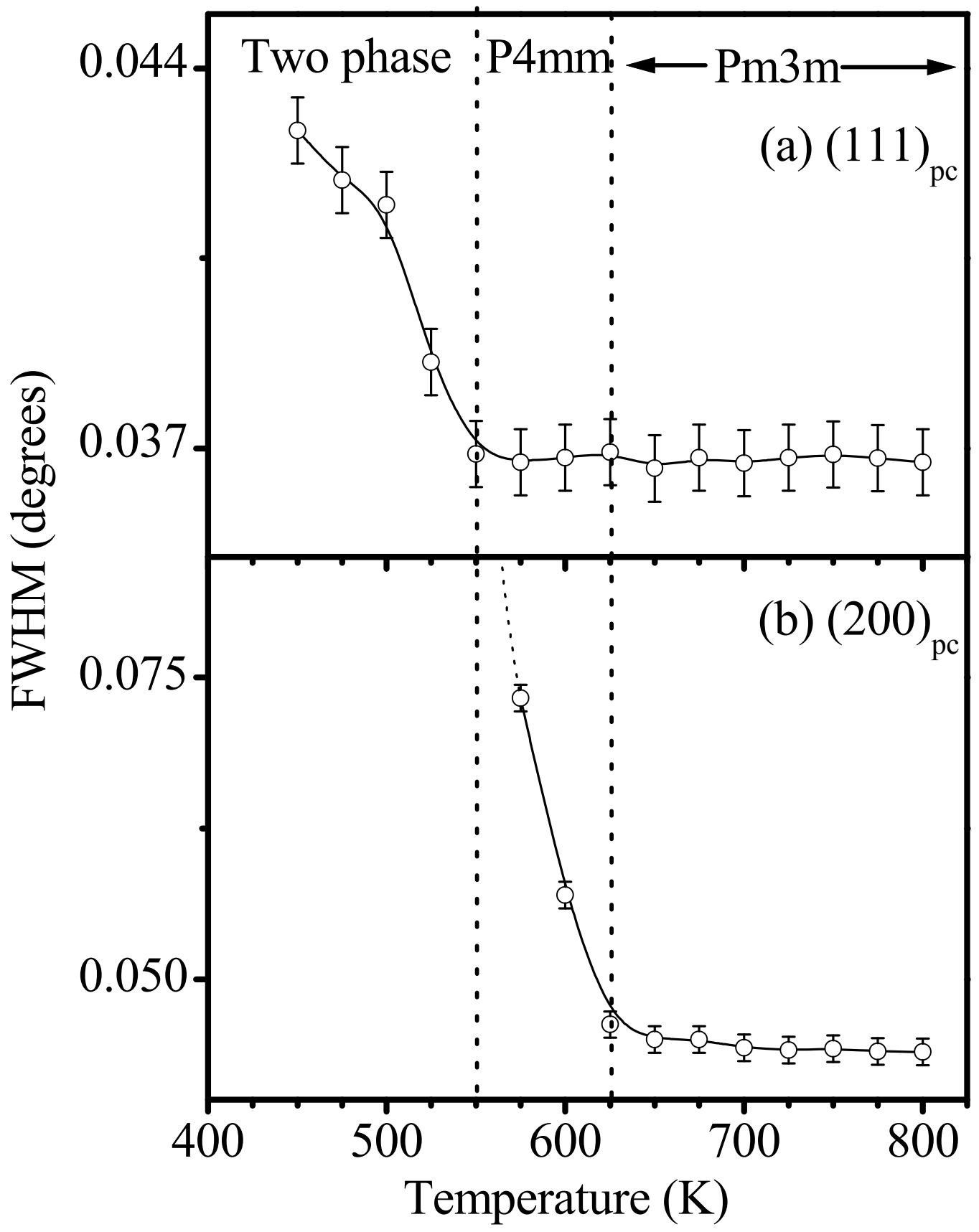

***Fig.6***

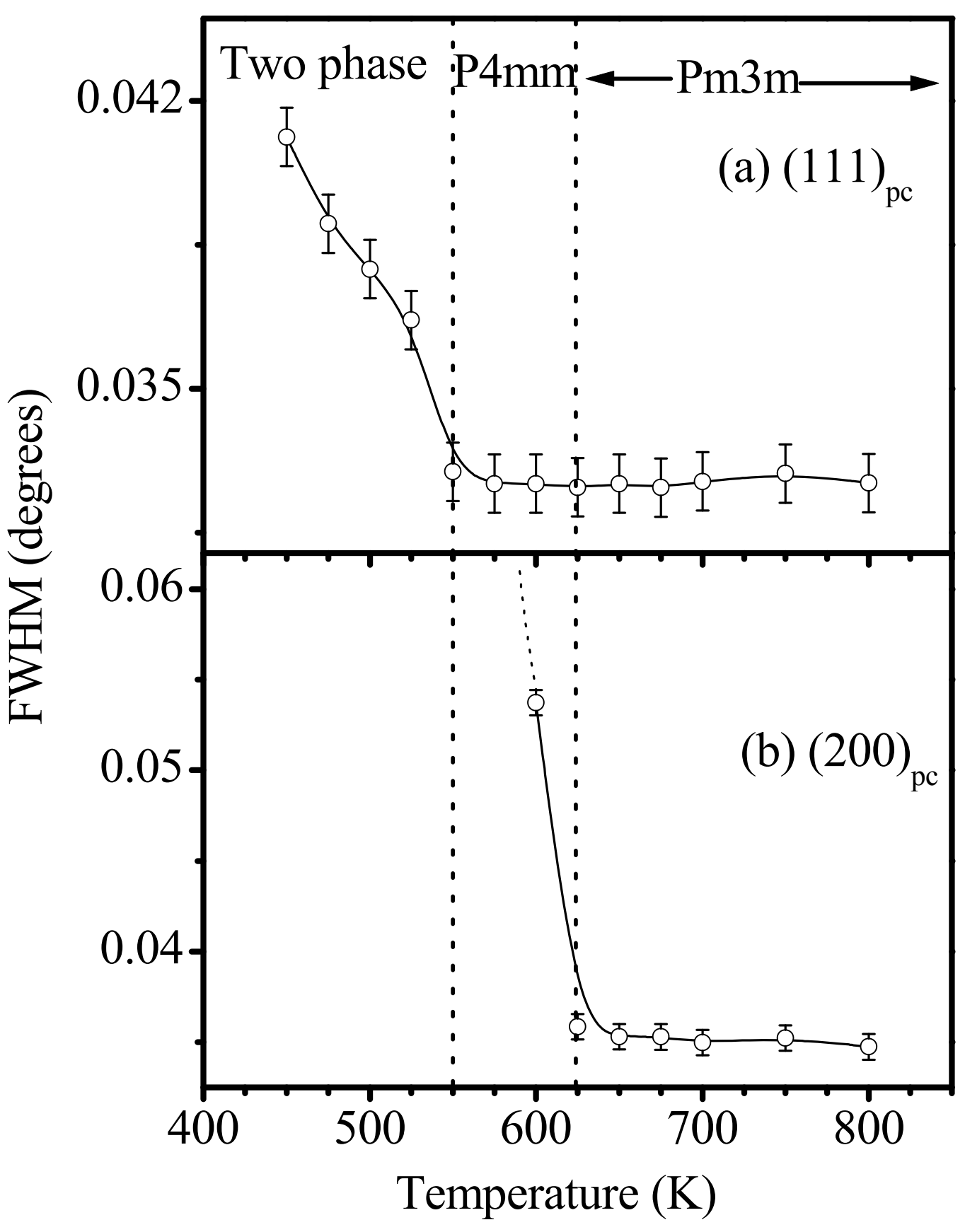

**Fig.7**

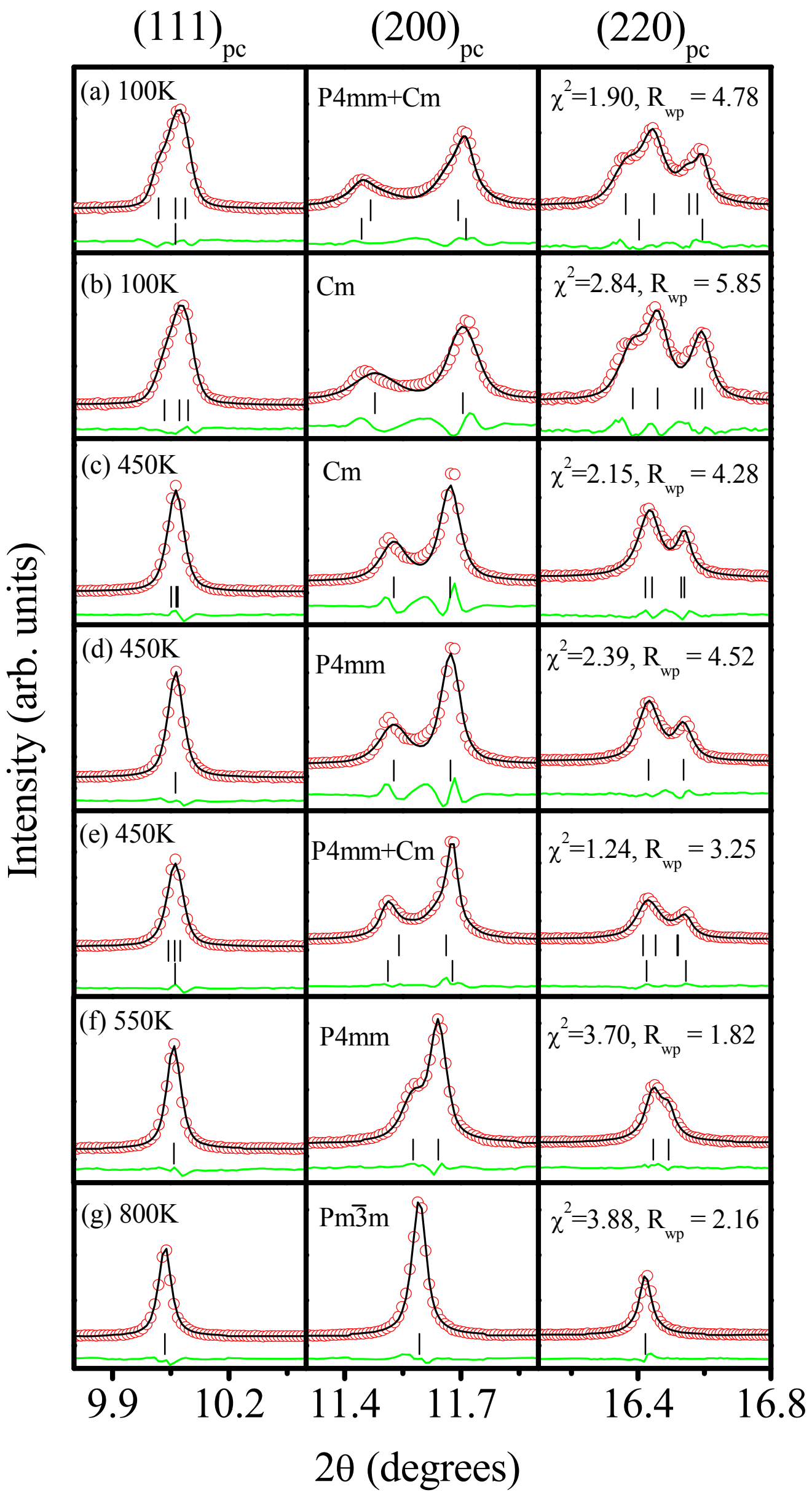

(111)pc
(200)pc
(220)pc
(a) 100K
P4mm+Cm
χ²=1.90, Rwp = 4.78
(b) 100K
Cm
χ²=2.84, Rwp = 5.85
(c) 450K
Cm
χ²=2.15, Rwp = 4.28
(d) 450K
P4mm
χ²=2.39, Rwp = 4.52
(e) 450K
P4mm+Cm
χ²=1.24, Rwp = 3.25
(f) 550K
P4mm
χ²=3.70, Rwp = 1.82
(g) 800K
Pm3̄m
χ²=3.88, Rwp = 2.16
Intensity (arb. units)
9.9
10.2
11.4
11.7
16.4
16.8
2θ (degrees)

**Fig.8**

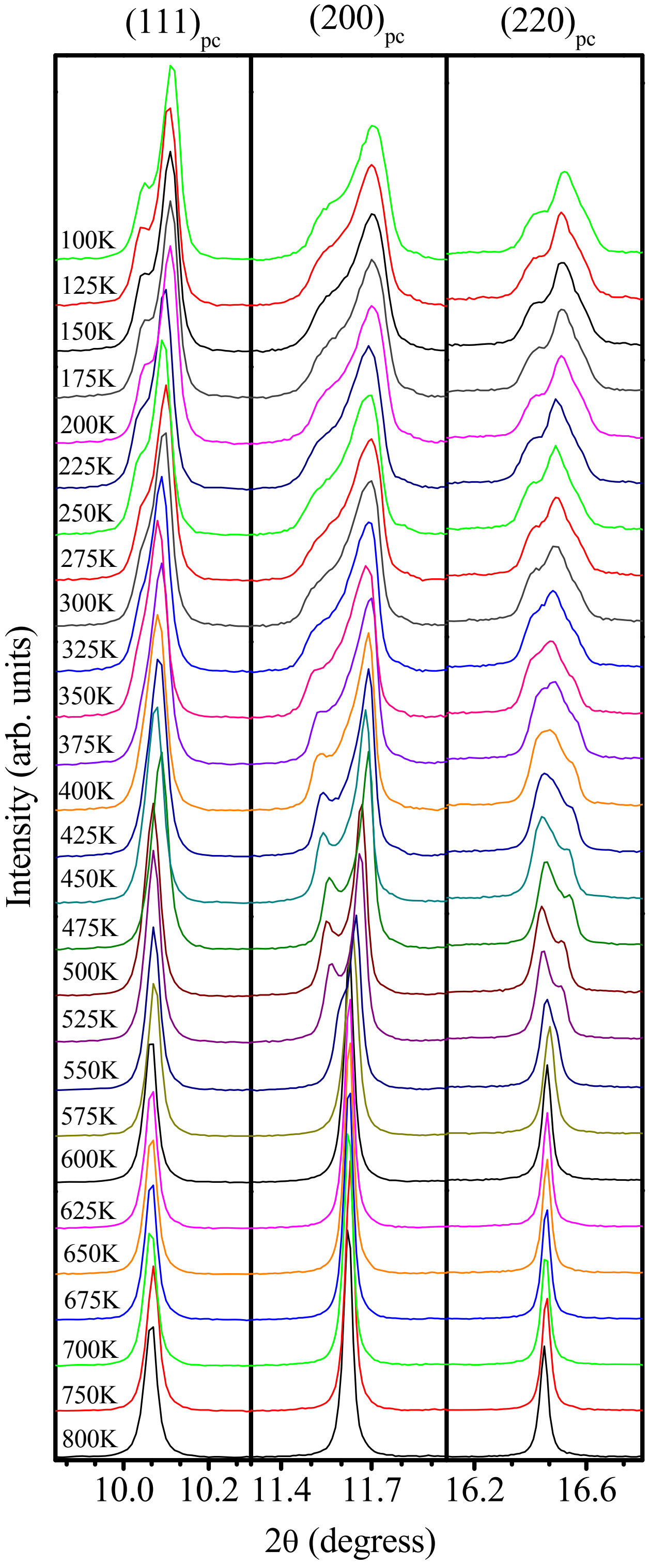

**Fig.9**

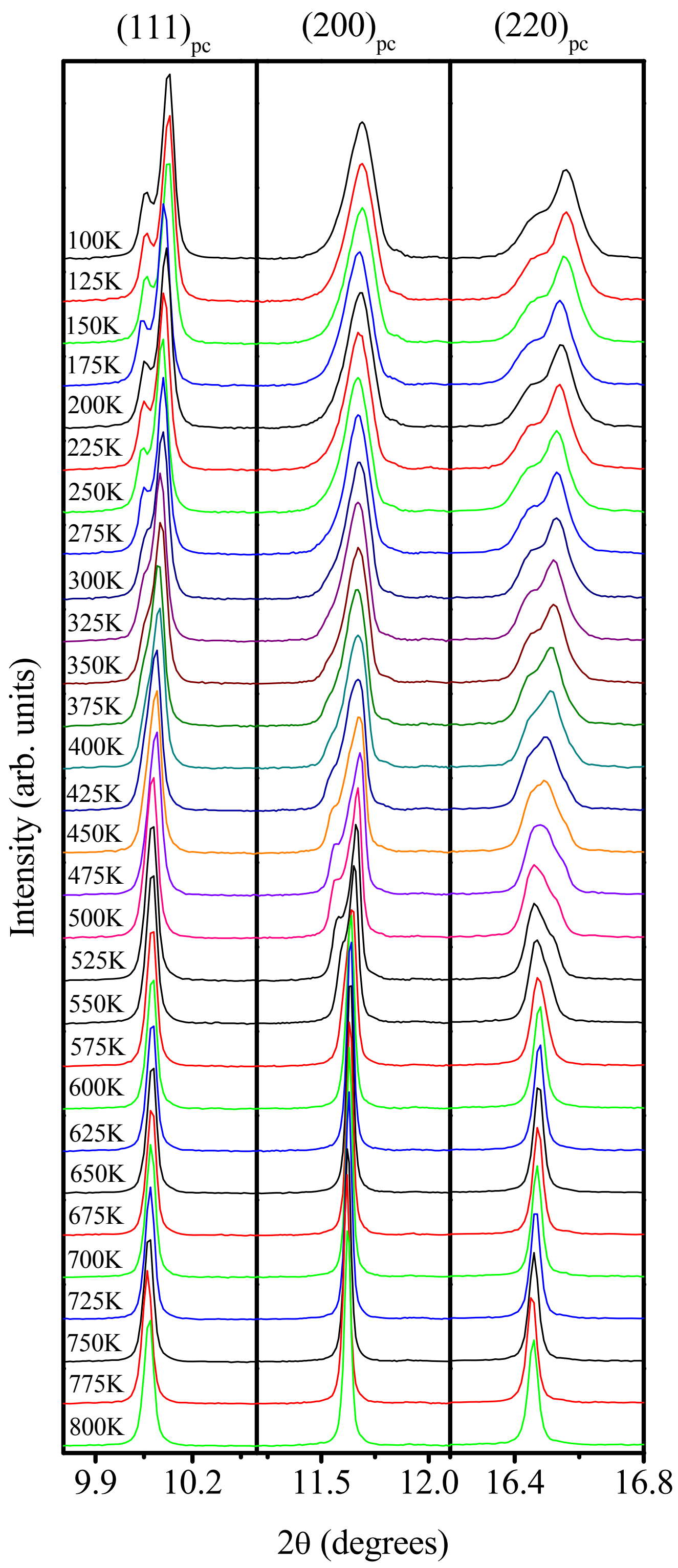

**Fig.10**

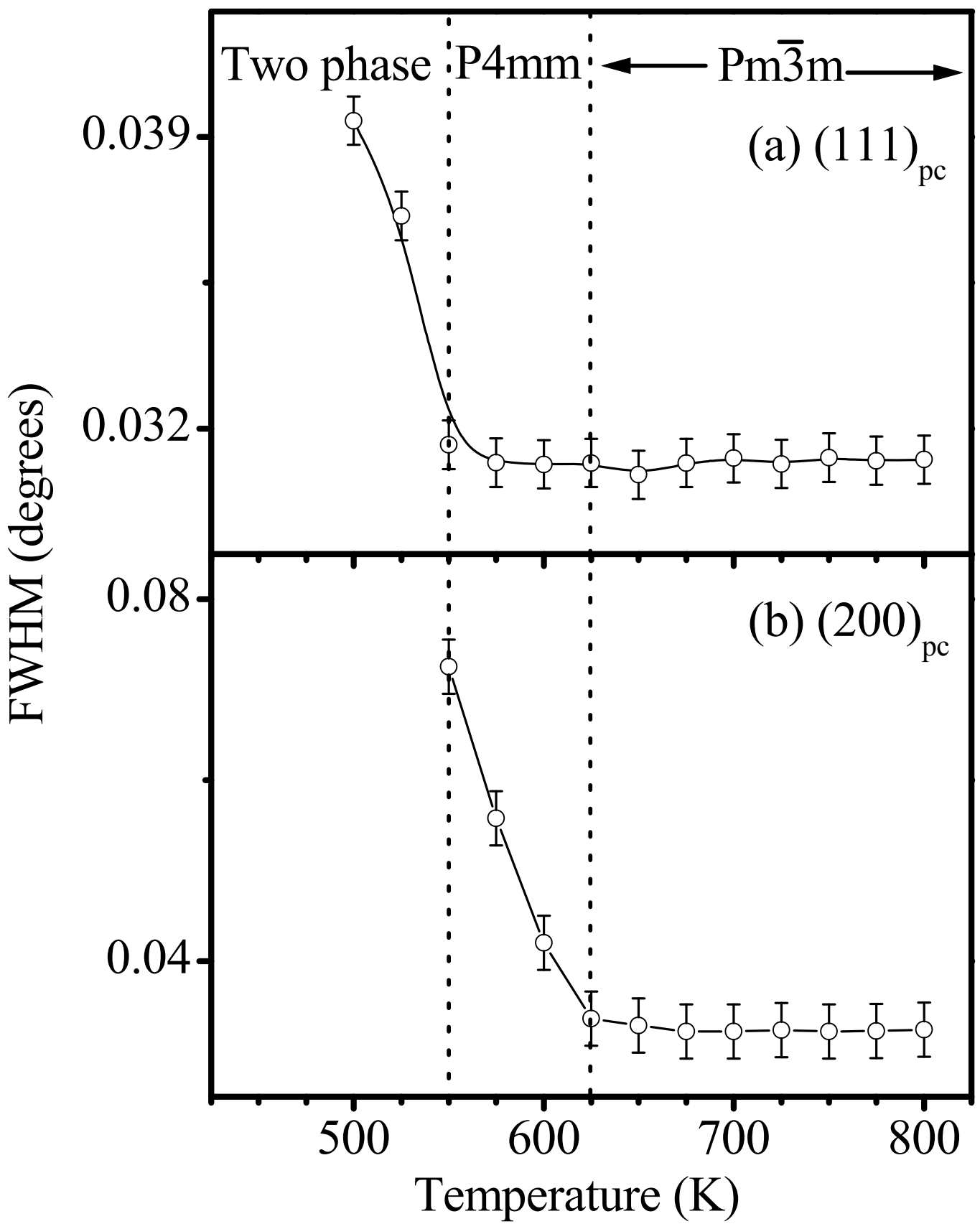

**Fig.11**

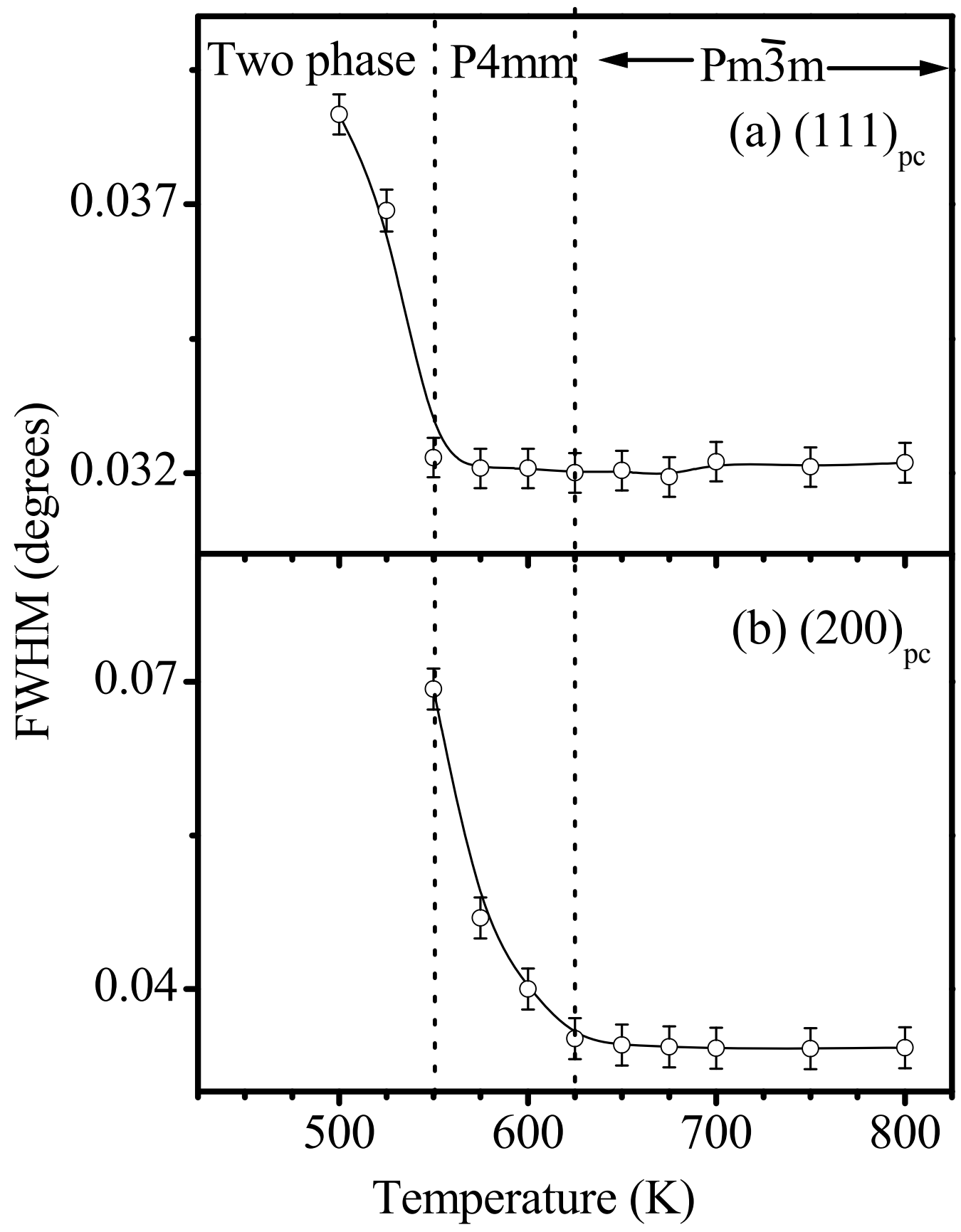

**Fig.12**

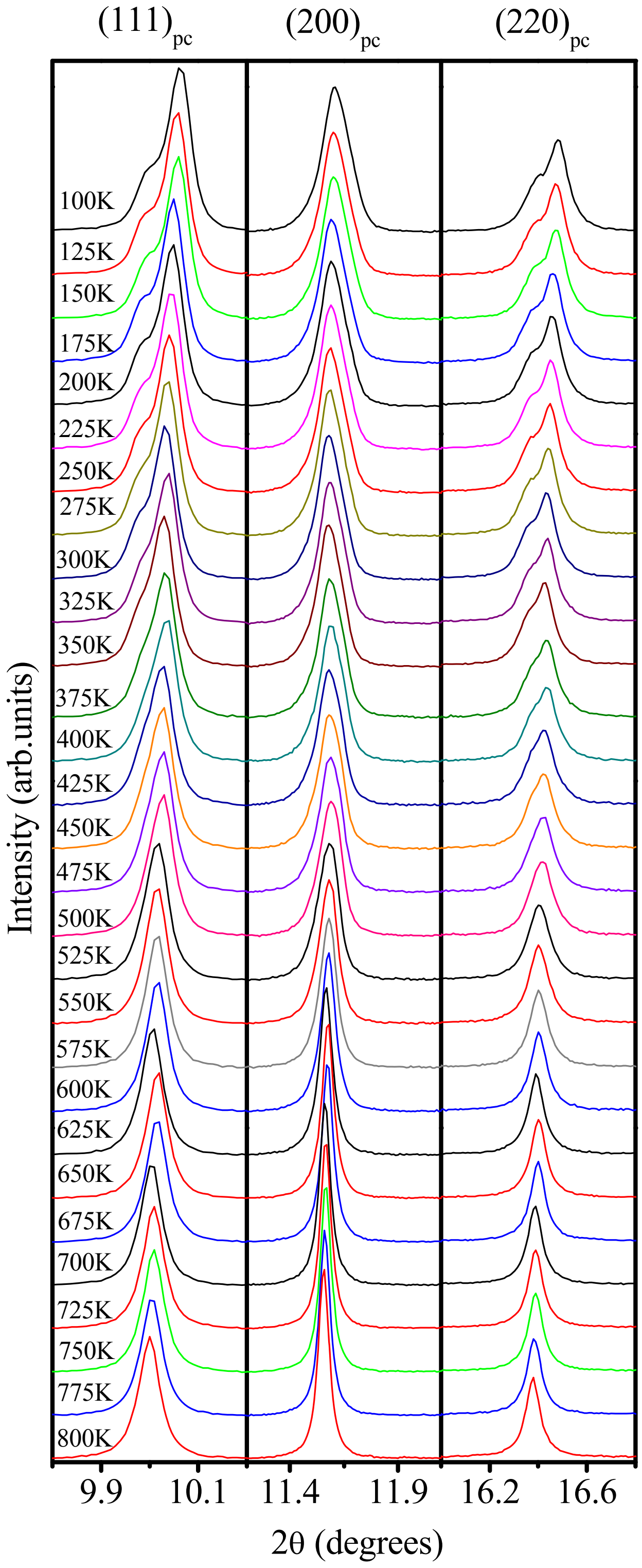

**Fig.13**

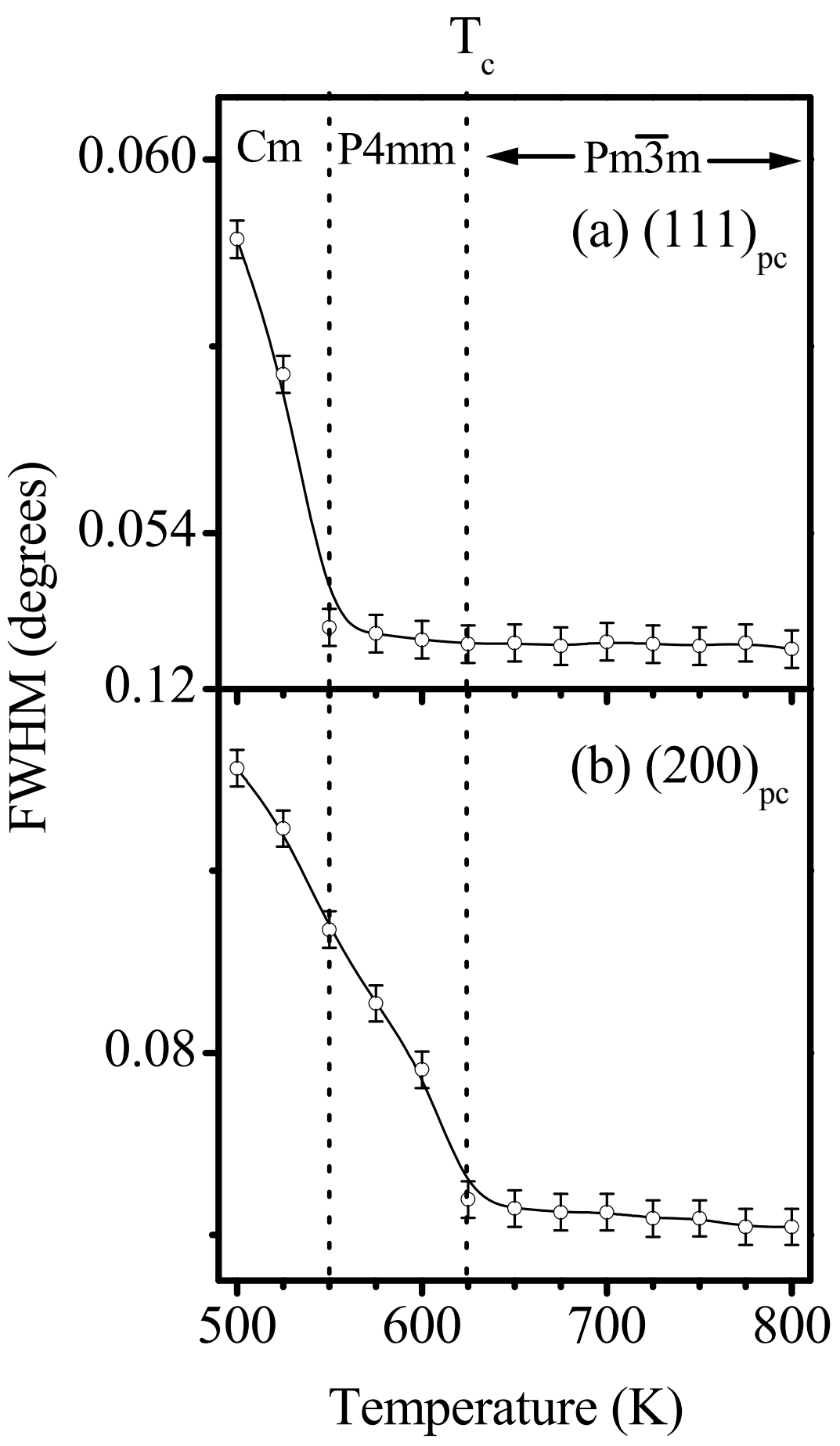

**Fig.14**

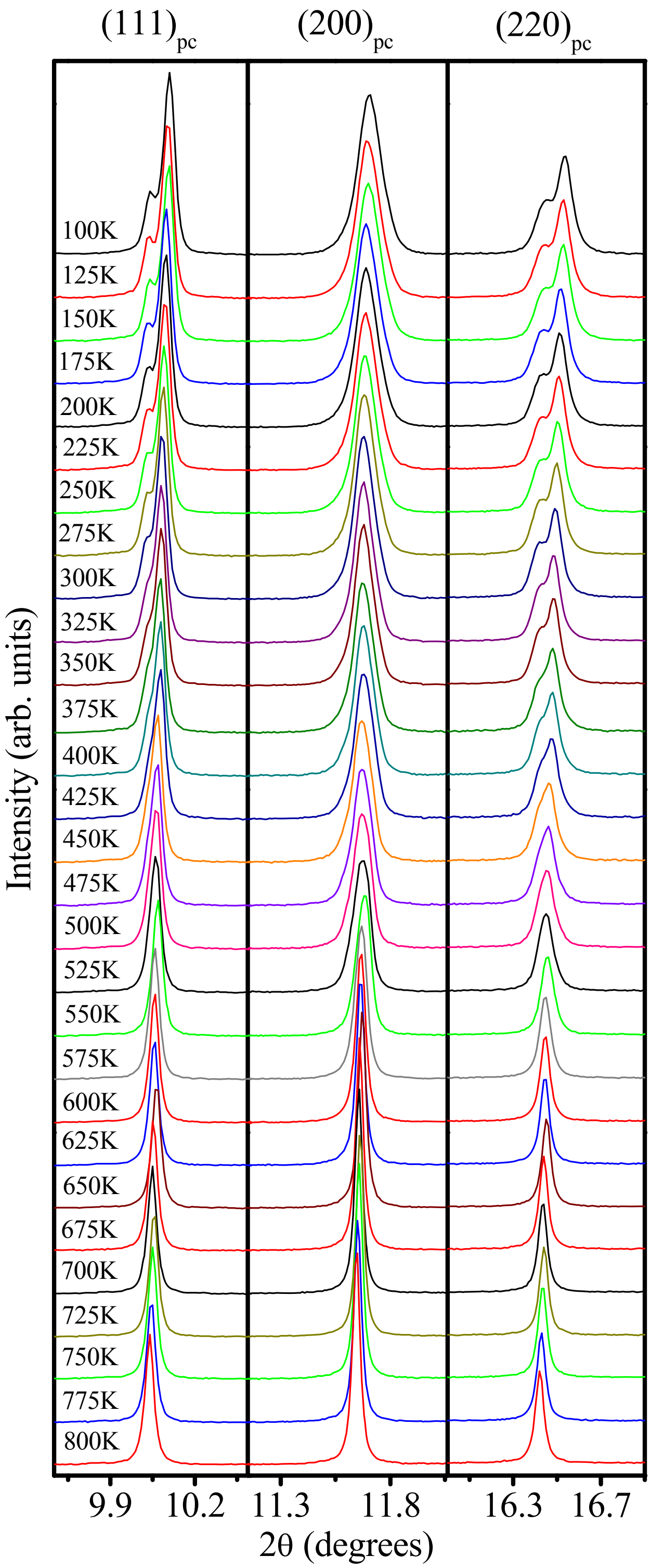

**Fig.15**

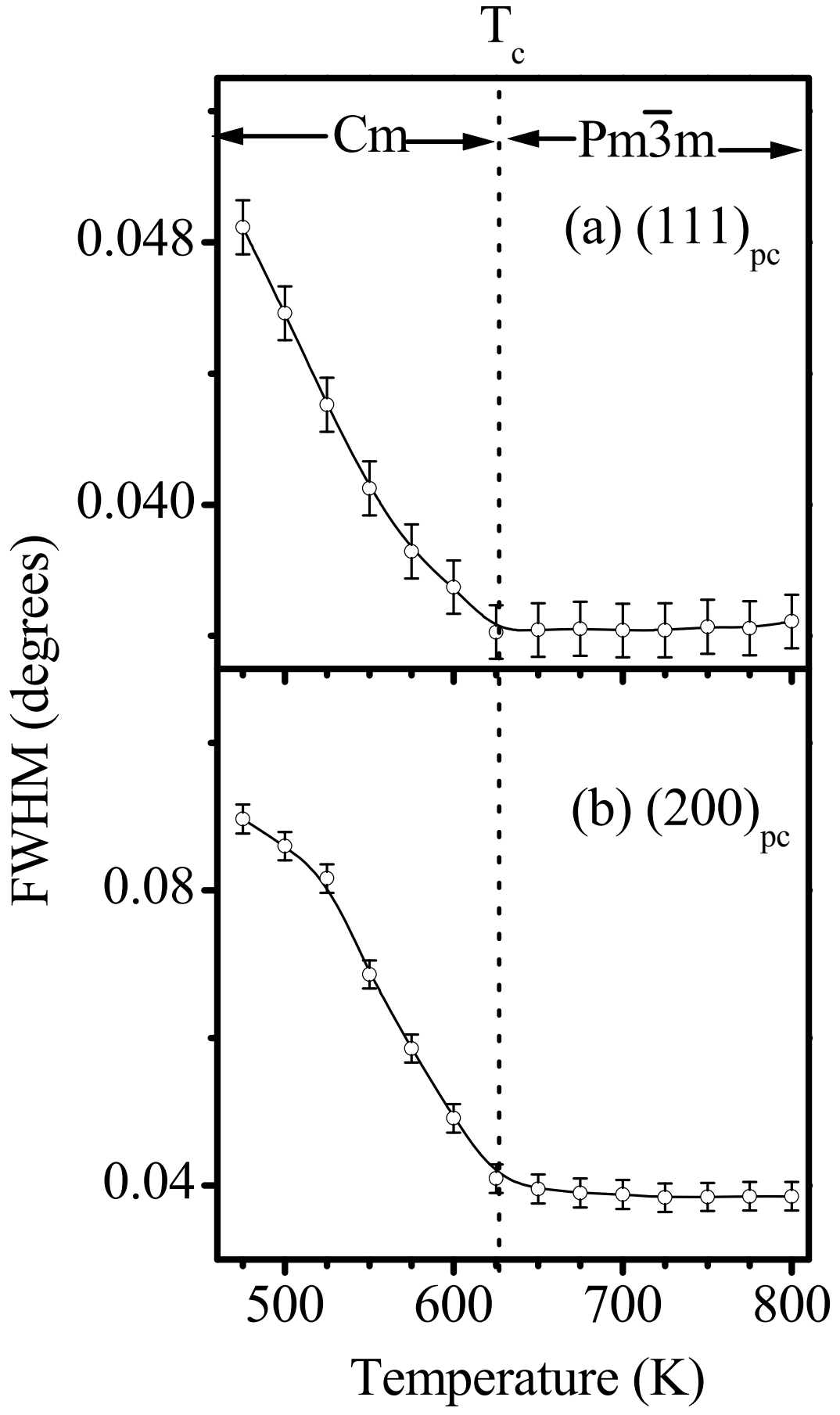

**Fig.16**

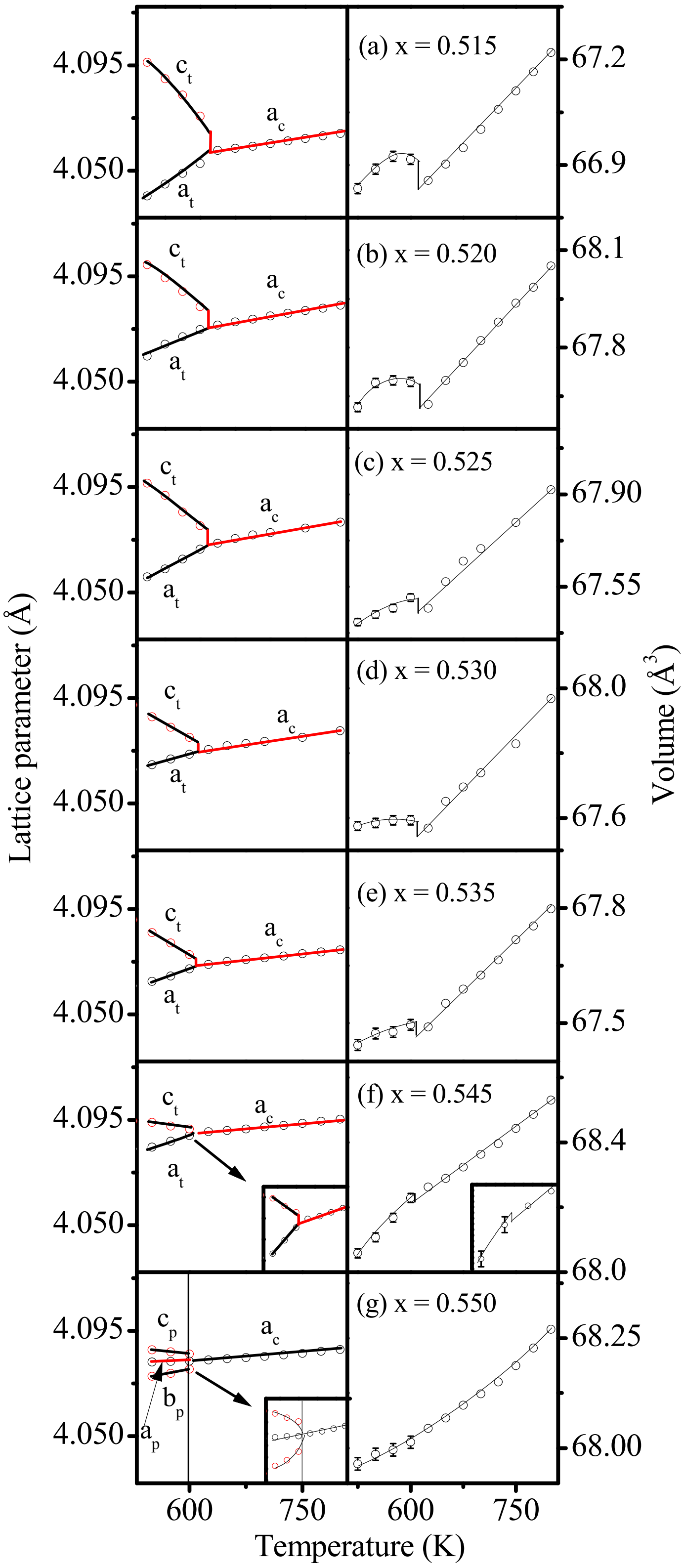

(a) x = 0.515
(b) x = 0.520
(c) x = 0.525
(d) x = 0.530
(e) x = 0.535
(f) x = 0.545
(g) x = 0.550
Lattice parameter (Å)
Volume (Å$^3$)
Temperature (K)

**Fig.17**

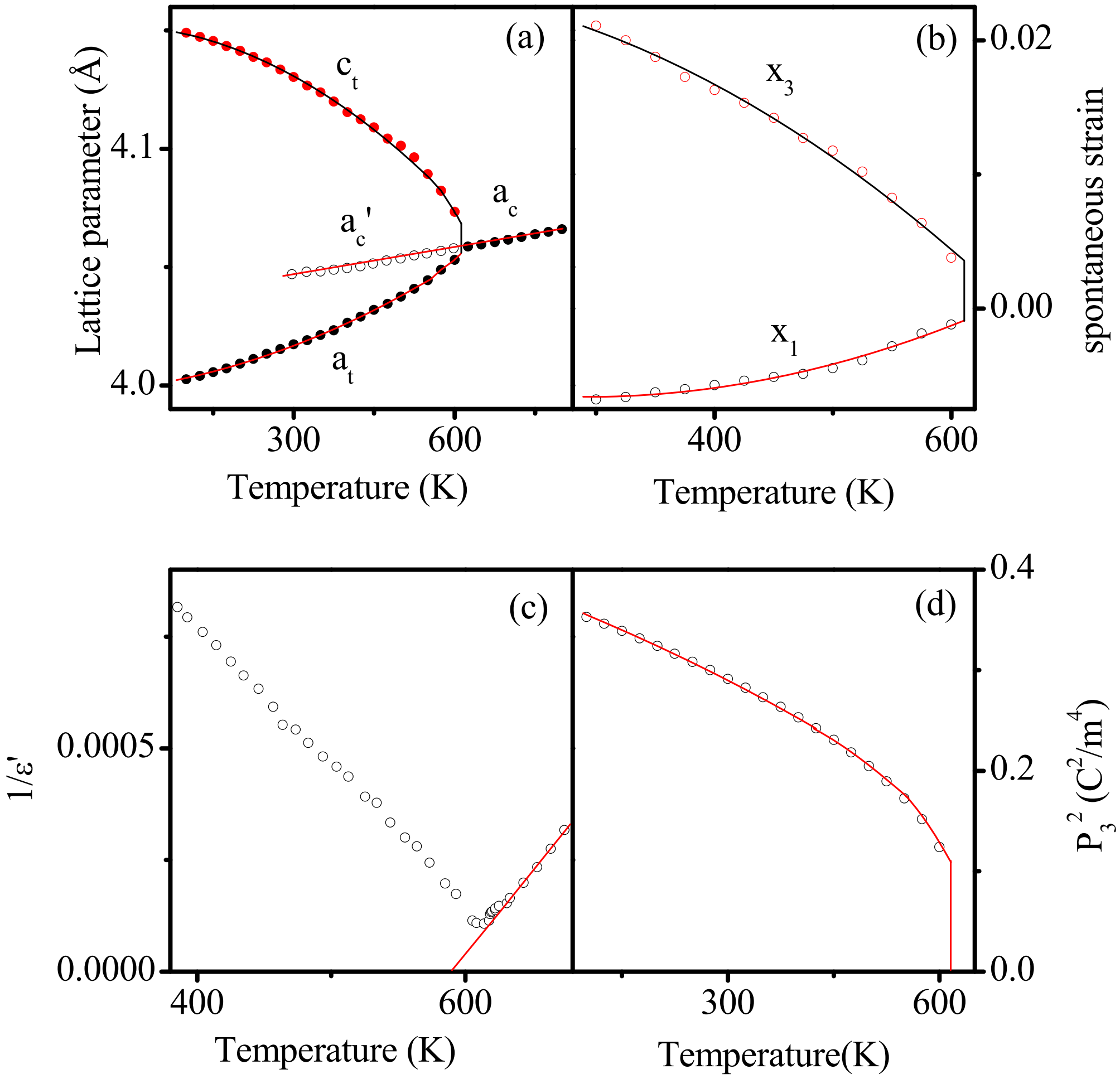

**Fig.18**

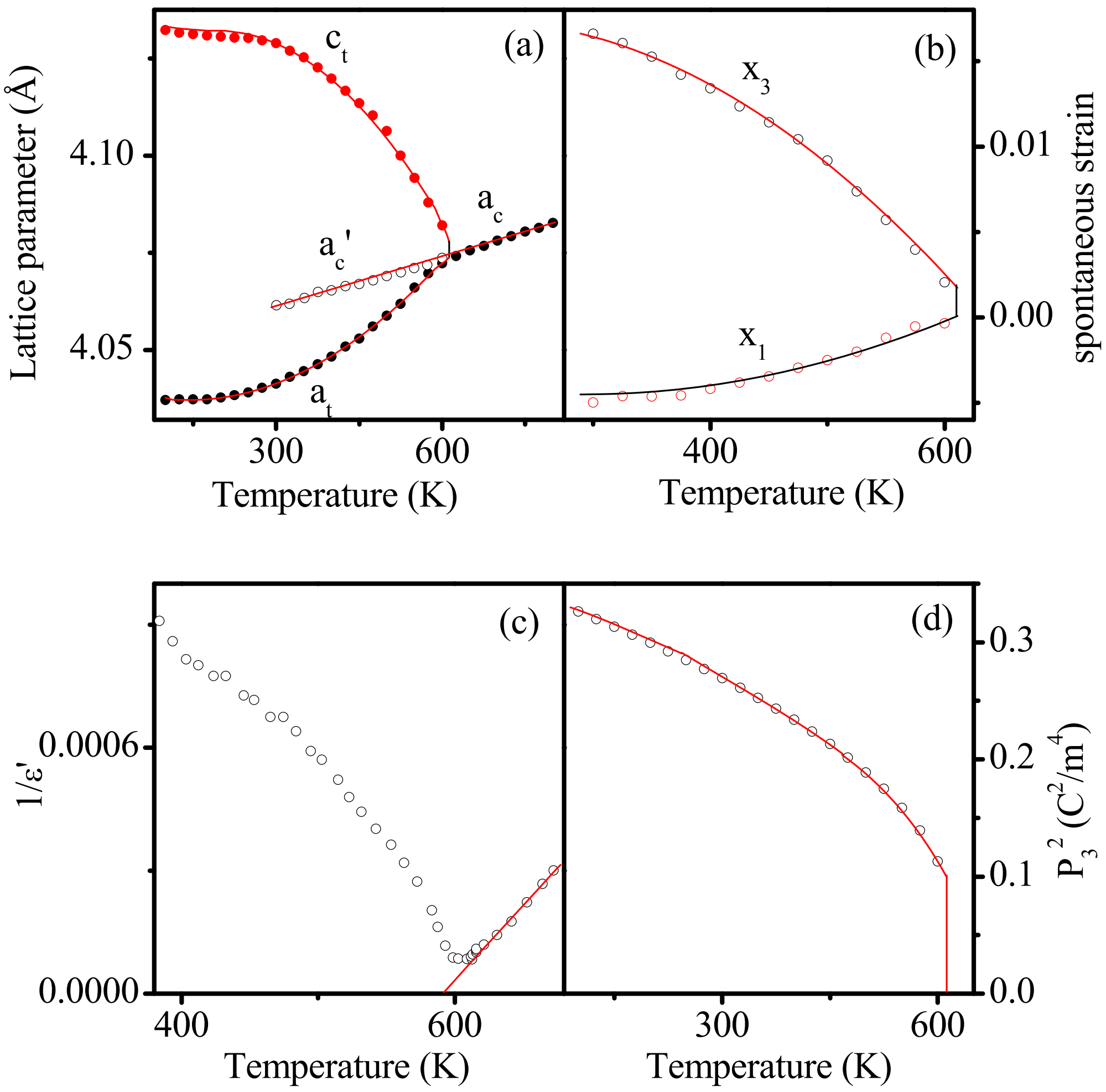

**Fig.19**

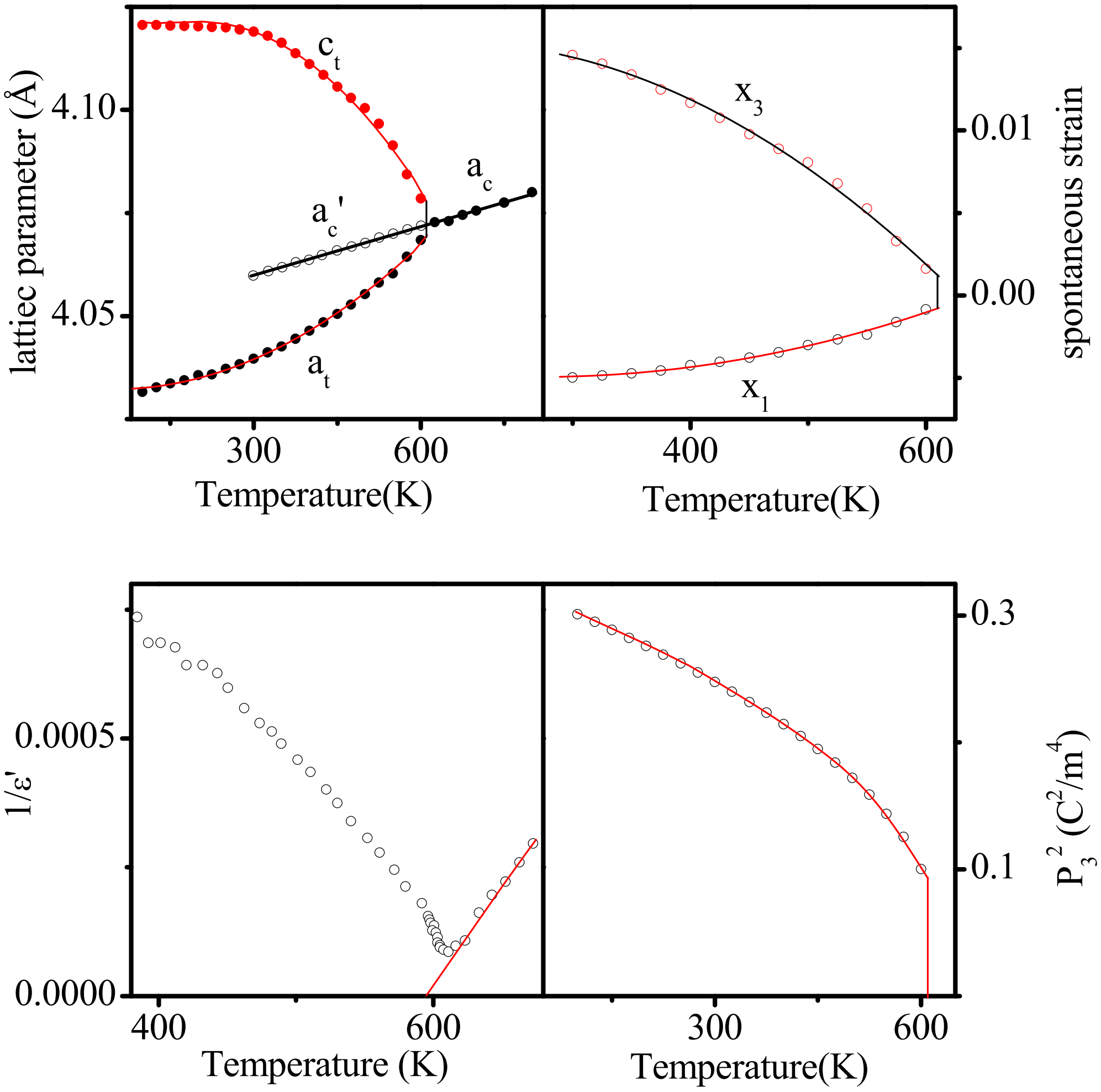

**Fig.20**

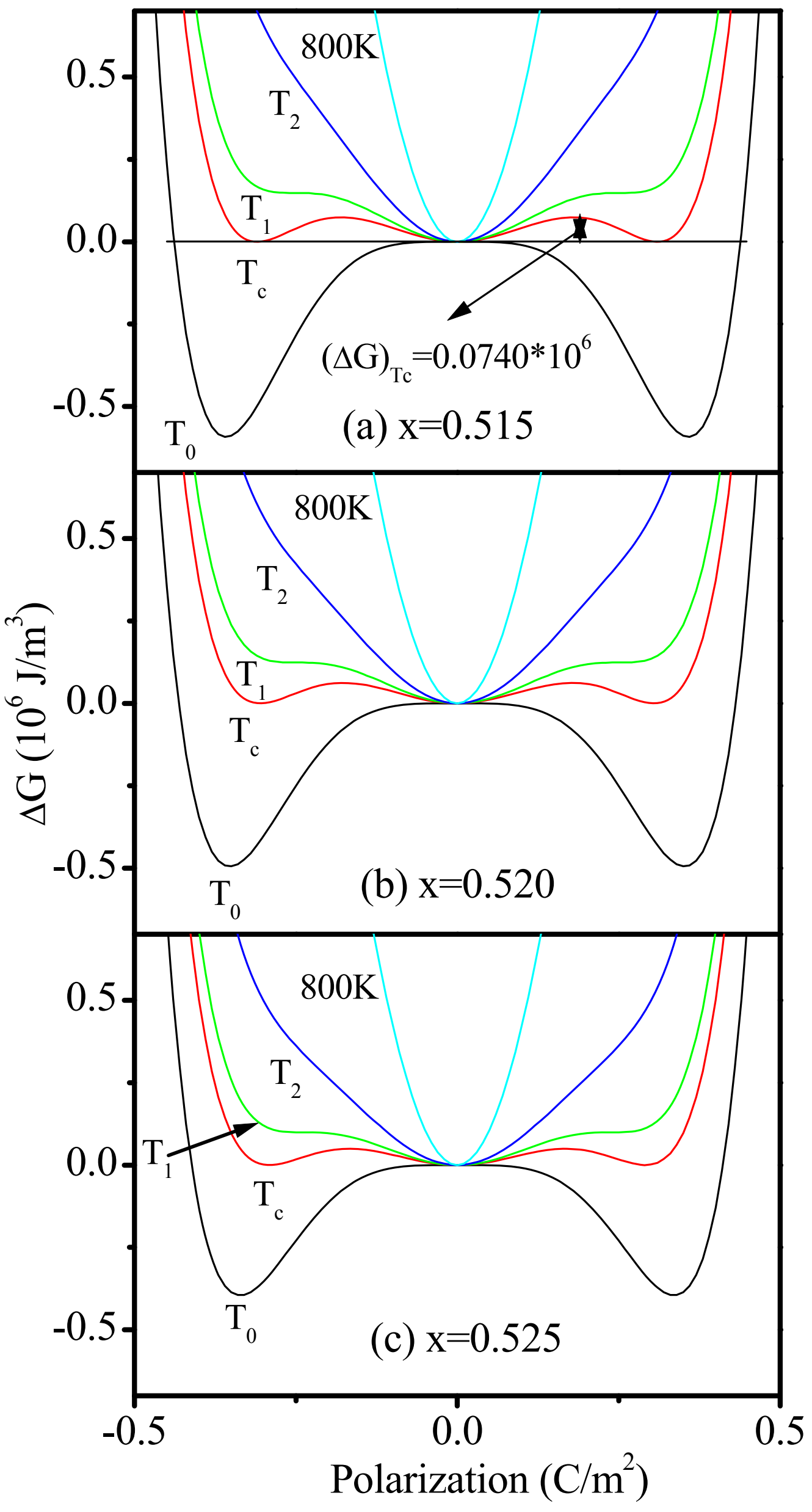

**Fig.21**

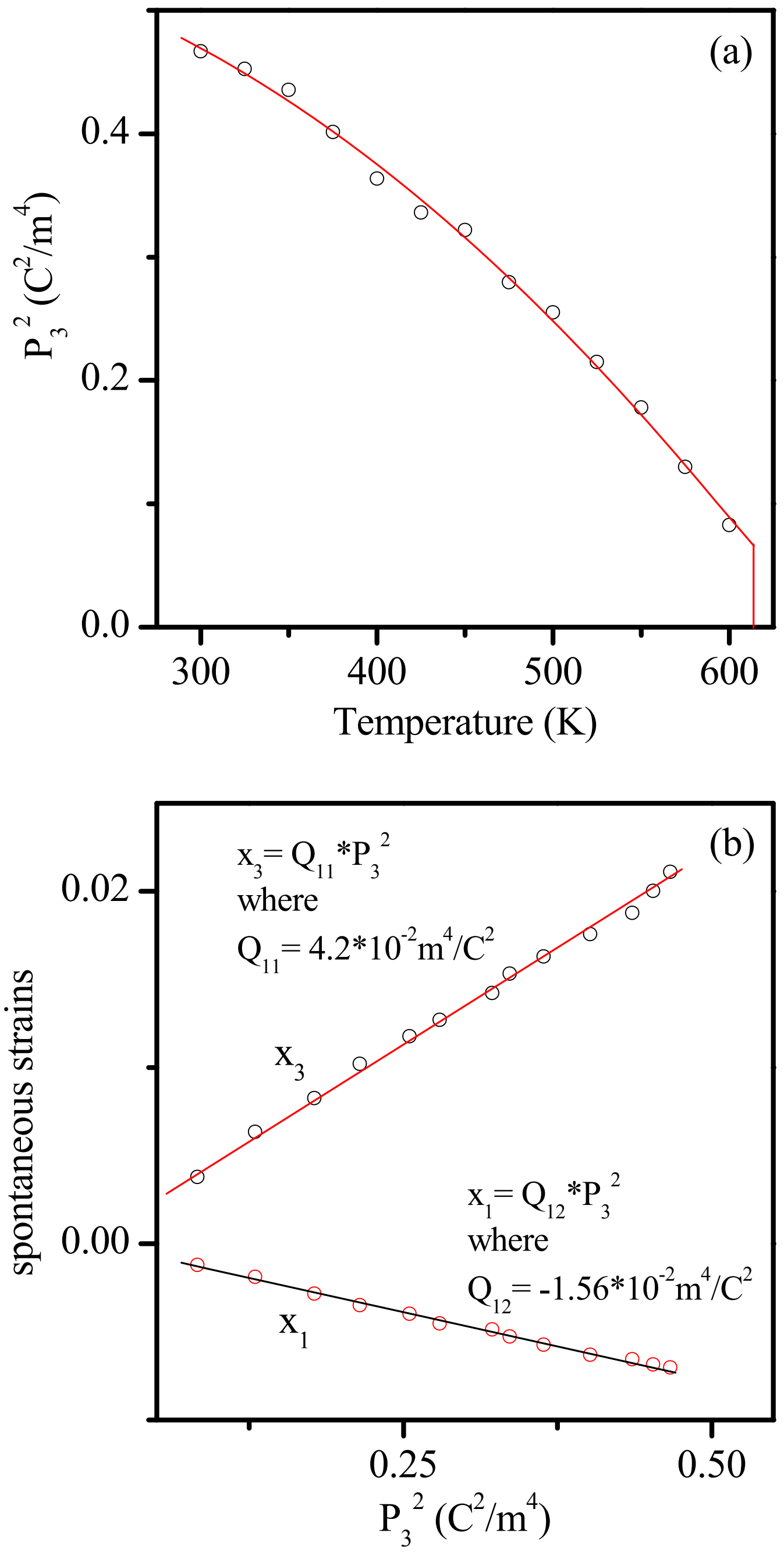

***Fig.22***

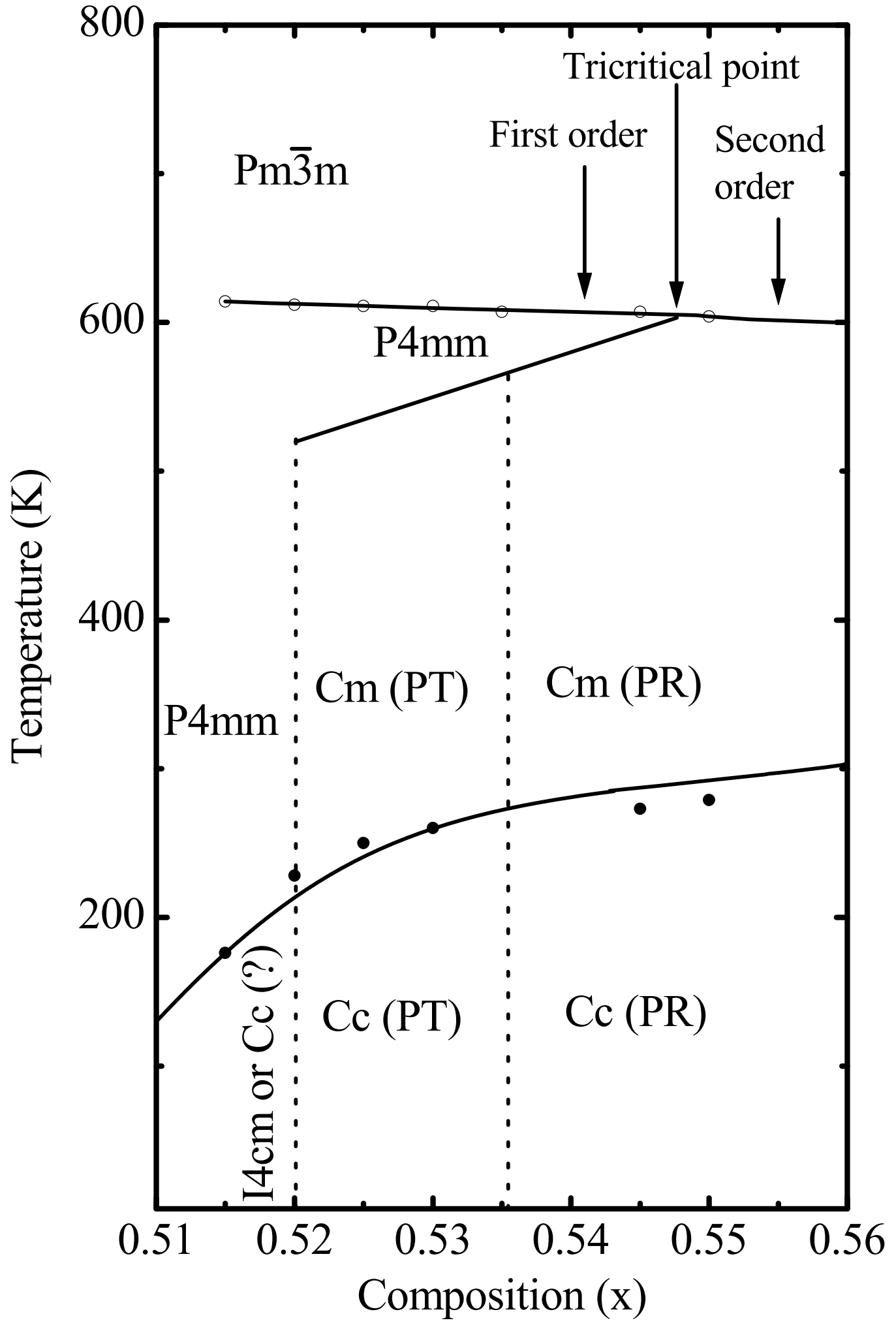

**Fig.23**

**Table I**

| **Composition (x)** | 0.515 | 0.520 | 0.525 |
|---|---|---|---|
| **$T_c$ (K)** | 614 | 612 | 611 |
| **$T_0$ (K)** | 589.353 | 590.1276 | 591.98 |
| **C ($10^5$K)** | 2.681 | 2.7576 | 2.726 |
| **$T_c$-$T_0$** | 24.647 | 21.8724 | 19.02 |
| **$x_{3c}$ ($10^{-4}$)** | 91.06 | 87.00 | 78.04 |
| **$Q_{11}$ ($10^{-2}m^4/C^2$)** | 9.468 | 9.355 | 9.224 |
| **$P_{3c}^2$($10^{-2}C^2/m^4$)** | 9.6176 | 9.30 | 8.46 |
| **$\alpha_{11}$ ($10^8m^5/C^2F$)** | -1.0801 | -0.9636 | -0.9319 |
| **$\alpha_{111}$ ($10^8m^9/C^4F$)** | 5.6151 | 5.1811 | 5.5077 |
| **$T_1$ (K)** | 622.2157 | 619.2908 | 617.3400 |
| **$T_2$ (K)** | 648.5058 | 642.6214 | 637.6280 |
| **$(\Delta G)_{Tc}$ ($10^6J/m^3$)** | 0.0740 | 0.0617 | 0.0494 |

**Table II**

| **Composition (x)** | 0.515 | 0.520 | 0.525 |
|---|---|---|---|
| **(i) PZT** | | | |
| $\alpha_{11}$ ($10^8 m^5/C^2F$) | -0.546816 | -0. 582517 | -0. 623225 |
| $\alpha_{111}$ ($10^8 m^9/C^4F$) | 1.33568 | 1.43996 | 1.50386 |
| **(ii) PSZT** (calculated using $Q_{11}$ of PZT) | | | |
| $\alpha_{11}$ ($10^8 m^5/C^2F$) | -1.0801 | -0.9636 | -0.9319 |
| $\alpha_{111}$ ($10^8 m^9/C^4F$) | 5.6151 | 5.1811 | 5.5077 |
| **(iii) PSZT** (calculated using $Q_{11}$ of PSZT515 ($Q_{11}$ was calculated using eqn. (9) and data given in Fig.22) | | | |
| $\alpha_{11}$ ($10^8 m^5/C^2F$) | -0.47912 | -0.43267 | -0.4243 |
| $\alpha_{111}$ ($10^8 m^9/C^4F$) | 1.1049 | 1.0444 | 1.1418 |